\documentclass[journal,twocolumn]{IEEEtran}
\usepackage{url}
\usepackage{algorithm}
\usepackage{algorithmicx}
\usepackage{graphicx}          
\graphicspath{{Figures/}}
\usepackage{mathrsfs}         
\usepackage{cite} 
\usepackage{amsmath}   
\usepackage{amsfonts}    
\usepackage{amssymb}   
\usepackage{booktabs}
\usepackage{threeparttable}
\usepackage{subfigure}
\usepackage{comment}
\usepackage{enumitem}
\usepackage{xcolor}
\usepackage{amsmath}
\usepackage{setspace}
\allowdisplaybreaks[4]
\usepackage[normalem]{ulem}
\newcommand\redsout{\bgroup\markoverwith{\textcolor{red}{\rule[0.5ex]{4pt}{2pt}}}\ULon}
\newtheorem{theorem}{\textbf{Theorem}}
\newtheorem{proposition}{\textbf{Proposition}}
\newtheorem{definition}{\textbf{Definition}}
\newtheorem{lemma}{\textbf{Lemma}}

\newtheorem{remark}{\textbf{Remark}}
\newtheorem{example}{\textbf{Example}}
\newtheorem{assumption}{\textbf{Assumption}}
\newtheorem{problem}{\textbf{Problem}}

\DeclareMathOperator{\diag}{diag}

\DeclareMathOperator{\uni}{Uni}

\newcommand*{\QEDA}{\hfill\ensuremath{\blacksquare}}   

\newcommand{\Cov}{\mathrm{Cov}}

\IEEEoverridecommandlockouts                              

\title{\LARGE \bf
	Asynchronous Distributed Reinforcement Learning for LQR Control via Zeroth-Order Block Coordinate Descent}

\author{Gangshan Jing,~He Bai,~Jemin George,~Aranya Chakrabortty~and~Piyush K. Sharma

\thanks{G.~Jing is with Chongqing University, Chongqing, 400044, PRC. 
{\tt\small jinggangshan@cqu.edu.cn}, A. Chkarabortty is with  North Carolina State University, Raleigh, NC 27695, USA. 
{\tt\small achakra2@ncsu.edu}, H.~Bai is with Oklahoma State University, Stillwater, OK 74078, USA. 
{\tt\small he.bai@okstate.edu}, J.~George and P.~Sharma are with the DEVCOM Army Research Laboratory, Adelphi, MD 20783, USA.
{\tt\small \{jemin.george.civ,piyush.k.sharma.civ\}@army.mil}}
}

\allowdisplaybreaks[4]

\begin{document}

	\maketitle

	\begin{abstract} 
		Recently introduced distributed zeroth-order optimization (ZOO) algorithms have shown their utility in distributed reinforcement learning (RL).  Unfortunately, in the gradient estimation process, almost all of them require random samples with the same dimension as the global variable and/or require evaluation of the global cost function, which may induce high estimation variance for large-scale networks. In this paper, we propose a novel distributed zeroth-order algorithm by leveraging the network structure inherent in the optimization objective, which allows each agent to estimate its local gradient by local cost evaluation independently, without use of any consensus protocol. The proposed algorithm exhibits an asynchronous update scheme, and is designed for stochastic non-convex optimization with a possibly non-convex feasible domain based on the block coordinate descent method. The algorithm is later employed as a distributed model-free RL algorithm for distributed linear quadratic regulator design, where a learning graph is designed to describe the required interaction relationship among agents in distributed learning. We provide an empirical validation of the proposed algorithm to benchmark its performance on convergence rate and variance against a centralized ZOO algorithm.
	\end{abstract}
	\vspace{-1.5em} 

	\section{Introduction}
	Zeroth-order optimization (ZOO) algorithms solve optimization problems with implicit function information by estimating gradients via zeroth-order observation (function evaluations) at judiciously chosen points \cite{Spall05}. They have been extensively employed as model-free reinforcement learning (RL) algorithms for black-box off-line optimization \cite{Nesterov17}, online optimization \cite{Flaxman04}, neural networks training \cite{Chen17}, and model-free linear quadratic regulator (LQR) problems \cite{Fazel18,malik2020derivative,Furieri20,Mohammadi20}. Although, ZOO algorithms are applicable to a broad range of problems, they suffer from high variance and slow convergence rate especially when the problem is of a large size. To improve the performance of ZOO, besides the classical one-point feedback \cite{Flaxman04,malik2020derivative}, other types of estimation schemes have been proposed, e.g., average of multiple one-point feedback \cite{Fazel18}, two-point feedback \cite{Agarwal10,Nesterov17,Chen17,Shamir17,malik2020derivative}, average of multiple two-point feedback \cite{Mohammadi20}, and one-point residual feedback \cite{zhang2022new}. These methods have been shown to reduce variance and increase convergence rate efficiently. However, all these methods are still based on the global optimization objective evaluation, which may cause large errors and high variance of gradient estimates when encountering a large-size problem.
	
	Multi-agent systems (MASs) are one of the most representative systems that induce large-size optimizations \cite{sharma2021survey}. In recent years, distributed zeroth-order convex and non-convex optimizations on multi-agent networks have been extensively studied, e.g., \cite{Sun19,Hajine19,Gratton20,Tang20,Akhavan21}, all of which decompose the original cost function into multiple functions assigned to the agents. Unfortunately, the individual variable dimension is the same as that of the original problem. As a result, the agents are expected to eventually reach consensus on the optimization variables. In such a setting, the high variance of the gradient estimate caused by the high variable dimension remains unchanged. In \cite{Li19}, each agent has a lower dimensional variable. However, still consensus was employed to estimate the global cost function value. Such a framework maintains a similar gradient estimate to that of the centralized ZOO algorithm, which still suffers a high variance when the network is large.  

    Block coordinate descent (BCD) is known to be efficient for large-size optimizations due to the low per-iteration cost \cite{nesterov2012efficiency,peng2016coordinate}. The underlying idea is to solve optimizations by updating only partial components (one block) of the entire variable in each iteration, \cite{canutescu2003cyclic,nesterov2012efficiency,richtarik2014iteration,wright2015coordinate,peng2016coordinate,xu2017globally}. Recently, zeroth-order BCD algorithms have been proposed to craft adversarial attacks for neural networks \cite{Chen17,cai2021zeroth}, where the convergence analysis was given under the convexity assumption on the objective function in \cite{cai2021zeroth}. A non-convex optimization was addressed by a zeroth-order BCD algorithm in \cite{yu2019zeroth}, whereas the feasible domain was assumed to be convex.

	In this paper, we consider a general stochastic locally smooth non-convex optimization with a possibly non-convex feasible domain, and propose a distributed ZOO algorithm based on BCD. Out of the aforementioned situations, we design local cost functions involving only partial agents by utilizing the network structure embedded in the overall optimization objective. This is reasonable in model-free control problems because although the dynamics of each agent is unknown, the inter-agent coupling structure may be known and the control objective with an inherent network structure is usually artificially designed. Our algorithm allows each agent to update its optimization variable by evaluating a local cost independently, without requiring consensus on agents' optimization variables. This formulation is applicable to distributed LQR control and many multi-agent cooperative control problems such as formation control \cite{Oh15}, network localization \cite{fang2020angle} and cooperative transportation \cite{ebel2018distributed}, where the variable of each agent converges to a desired local set corresponding to a part of the minimizer of the global cost function. Our main contributions are listed as follows.

    (i). We propose a distributed accelerated zeroth-order algorithm with asynchronous sample and update schemes (Algorithm \ref{alg:as}). The MAS is divided into multiple clusters, where agents in different clusters evaluate their local costs asynchronously, and the evaluations for different agents are completely independent. It is shown that with appropriate parameters, the algorithm converges to an approximate stationary point of the optimization with high probability. The sample complexity is polynomial with the reciprocal of the convergence error, and is dominated by the number of clusters (see Theorem \ref{th as}). When compared to ZOO based on the global cost evaluation, our algorithm produces gradient estimates with lower variance and thus results in faster convergence (see the variance analysis in Subsection \ref{subsec: as variance}).
		
	(ii). We further consider a model-free multi-agent distributed LQR problem, where multiple agents with decoupled dynamics\footnote{Our algorithm applies to MASs with coupled dynamics. Please refer to Remark \ref{re coupled dynamics} for more details.} cooperatively minimize a cost function involving all the agents. The optimal LQR controller is desired to be distributed, i.e., the control of each agent involves only its neighbors. We describe the couplings in the cost function and the structural constraint by a \emph{cost graph} and a \emph{sensing graph}, respectively. To achieve distributed learning, we design a local cost for each agent, and introduce a \emph{learning graph} that describes the required agent-to-agent interaction relationship for local cost evaluations. Specifically, each local cost is designed such that its gradient w.r.t. an individual control gain is the same as that of the global cost. The learning graph is determined by the cost and sensing graphs and is needed only during the learning stage.  By implementing our algorithm, the agents optimize their local costs and learn distributed controllers corresponding to a stationary point of the optimization distributively. The design of local costs and the learning graph plays the key role in enabling distributed learning. The learning graph is typically denser than the cost and sensing graphs, which is a trade-off for not using a consensus algorithm.
	
	The comparisons between our work and existing related results are listed as follows.
 
	(i). In comparison to centralized ZOO \cite{Flaxman04,Agarwal10,Nesterov17,Chen17,zhang2022new}, we reduce the variable dimension for each agent, and construct local costs involving only partial agents. Compared with distributed ZOO \cite{Sun19,Hajine19,Gratton20,Tang20,Akhavan21} and consensus-based distributed optimization \cite{swenson2021distributed,ren2021distributed}, our framework avoids the influence of the convergence error and convergence rate of the consensus algorithm, and results in reduced variance and high scalability to large-scale networks. Moreover, since there is no need to reach agreement on any global index, our algorithm benefits for privacy preserving. 

	(ii). In comparison to ZOO algorithms for LQR, e.g., \cite{Fazel18,malik2020derivative,Mohammadi20}, our problem has a structural constraint on the desired control gain, thus the cost function is not gradient dominated.
		
	(iii) Existing BCD algorithms in \cite{canutescu2003cyclic,nesterov2012efficiency,richtarik2014iteration,wright2015coordinate,peng2016coordinate,xu2017globally} always require global smoothness of the objective function. When only zeroth-order information is available, existing BCD algorithms require convexity of either the objective function \cite{cai2021zeroth} or the feasible domain \cite{yu2019zeroth}. In contrast, we propose a zeroth-order BCD algorithm for a stochastic non-convex optimization with a locally smooth objective and a possibly non-convex feasible set. A clustering strategy and a cluster-wise update scheme are proposed as well, which further improve the convergence rate.
		
	(iv) The distributed LQR problem is challenging because both the objective function and the feasible set are non-convex, and the number of locally optimal solutions grow exponentially in system dimension \cite{Feng19,Bu19to}. Existing results for distributed LQR include model-based centralized approaches \cite{Borrelli08,Bu19}, model-free centralized approaches \cite{Furieri20,Jing21}, and model-free distributed approaches \cite{Li19,Jingtcns}. All of them focus on finding a sub-optimal distributed controller for an approximate problem or converging to a stationary point of the optimization via policy gradient. Our algorithm is a derivative-free distributed policy gradient algorithm, which also seeks a stationary point of the problem and yields a sub-optimal solution. In \cite{alonso2022data,talebi2021distributed,yu2023online}, model-free algorithms have also been proposed to seek structured controllers. However, some restrictive assumptions have been made, such as decoupled objective functions \cite{alonso2022data}, homogeneous dynamics, identical structures for the cost function and the distributed controller\cite{talebi2021distributed}, and absence of a structured cost function \cite{yu2023online}, making them inapplicable to the problem studied in this paper.

	(v). The sample complexity of our algorithm is higher than that in \cite{Fazel18,malik2020derivative,Mohammadi20} because their optimization problems have the gradient domination property, and \cite{Fazel18,malik2020derivative} assume that the infinite horizon LQR cost can be obtained perfectly. The sample complexity of our algorithm for LQR is slightly lower than that in \cite{Li19} (see the discussion after Theorem \ref{th as LQR}). However, two consensus algorithms are employed in \cite{Li19} for distributed sampling and global cost estimation, respectively, which may slow down the gradient estimation process especially for large-scale problems. Moreover, the gradient estimation in \cite{Li19} is essentially based on global cost evaluation, while we adopt local cost evaluation, which improves the scalability.
		
	(vi). Distributed RL has also been extensively studied via a Markov decision process (MDP) formulation, e.g.,  \cite{kar2013cal,omidshafiei2017deep,lowe2017multi,Zhang2018fully,chen2021communication}. However, in comparison to our work, they require more information for each agent or are essentially based on global cost evaluation. More specifically, in  \cite{kar2013cal,lowe2017multi,Zhang2018fully,chen2021communication}, the global state is assumed to be available for all the agents. It has been shown in \cite{lowe2017multi} that naively applying policy gradient methods to multi-agent settings exhibits high variance gradient estimates. \cite{omidshafiei2017deep} describes a distributed learning framework with partial observations, but only from an empirical perspective. 
	
This paper is structured as follows. Section \ref{sec:problem} describes the optimization problem. Section \ref{sec: as} presents our zeroth-order BCD algorithm with convergence analysis. Section \ref{sec MAS} introduces the application of our algorithm to the model-free multi-agent LQR problem. Section \ref{sec: sim} shows asimulation example. Section \ref{sec: con} concludes the whole paper.

	\textbf{Notations.} Given function $f(x)$ with $x=(x_1^{\top},...,x_N^{\top})^{\top}$, denote $f(y_i,x_{-i})=f(x')$ with $x'=({x'}_1^{\top},...,{x'}_N^{\top})^\top$, $x'_i=y_i$ and $x'_j=x_j$ for all $j\neq i$. Let $\mathbb{R}^{n}$ and $\mathbb{R}_{\geq0}$ denote the $n-$dimensional Euclidean space and the space of nonnegative real numbers. The norm $\|\cdot\|$ denotes the  $l$2-norm for vectors, and denotes the spectral norm for matrices, $\|\cdot\|_F$ is the Frobenius norm. The symbol $\mathbb{E}[\cdot]$ denotes expectation, $\Cov(x)$ denotes the covariance matrix of vector $x$. Given a set $S$, $\text{Uni}(S)$ represents the uniform distribution in $S$, and $\partial S$ denotes the boundary of $S$. Given matrix $X=[X_{ij}]\in\mathbb{R}^{m\times n}$, $\text{vec}(X)=(X_{11}...,X_{1n},X_{21},...,X_{2n},...,X_{m1},...,X_{mn})^\top\in\mathbb{R}^{mn}$ means the vectorization of $X$. On the contrary, $\text{vec}^{-1}$ is the inverse of the vectorization operator, i.e., $\text{vec}^{-1}(\text{vec}(X))=X$. Given another matrix $Y$, let $\langle X,Y\rangle=\text{trace}(X^{\top}Y)$ be the Frobenius inner product. The pair $\mathcal{G}=(\mathcal{V},\mathcal{E})$ denotes a directed or an undirected graph, where $\mathcal{V}$ is the set of vertices, $\mathcal{E}\subset\mathcal{V}^2$ is the set of edges. A pair $(i,j)\in\mathcal{E}$ implies that there exists an edge from $i$ to $j$. A path from vertex $i$ to vertex $j$ in $\mathcal{G}$ is a sequence of pairs $(i,i_1)$, $(i_1,i_2)$, ..., $(i_s,j)\in\mathcal{E}$.

	\section{Stochastic ZOO via Multi-Agent Networks}\label{sec:problem}
	In this section, we review the zeroth-order stochastic non-convex optimization problem and describe the optimization we aim to solve in this paper.
\vspace{-0.5cm}	
	\subsection{Stochastic ZOO}
	Consider a general optimization problem
	\begin{equation}\label{optimization}
		\mathop{\text{minimize}}\limits_{x\in\mathcal{X}}f(x),
	\end{equation}
	where \begin{equation}
	f(x)=\mathbb{E}_{\xi\sim\mathcal{D}}[h(x,\xi)]
	\end{equation}
	is continuously differentiable, $x\in\mathbb{R}^q$ is the optimization variable, $\mathcal{X}\subseteq\mathbb{R}^q$ is its feasible domain, unbounded, and possibly non-convex, $\xi\in\mathbb{R}^p$ is a random variable and may denote the noisy data with distribution $\mathcal{D}$ in real applications, and $h(\cdot,\cdot): \mathbb{R}^q\times\mathbb{R}^{p}\rightarrow\mathbb{R}_{\geq0}$ gives the observed cost.

\begin{assumption}\label{assumption1}
We make the following assumptions: 
\begin{enumerate}[label={\textbf{\Alph*:}}, ref={assumption~\Alph*}]
\item \label{assump-A}%
    The function $f$ is $(\lambda_x,\zeta_x)$ locally  Lipschitz\footnote{In practice, usually the local Lipschitz continuity property of $f(x)$ is obtained by local Lipschitz continuity of $h(x,\xi_0)$ with a fixed $\xi_0\in\mathbb{R}^p$ and boundedness of $\xi$'s moments.} in $\mathcal{X}$, i.e., for any $x\in\mathcal{X}$, if $\|x'-x\|\leq\zeta_x$ for $x'\in\mathbb{R}^q$, then
		\begin{equation}
			\|f(x')-f(x)\|\leq\lambda_x\|x'-x\|,
		\end{equation}
		where $\lambda_x$ and $\zeta_x$ are both continuous in $x$.
\item \label{assump-B}%
    The function $f$ has a $(\phi_x,\beta_x)$ locally Lipschitz gradient in $\mathcal{X}$, i.e., for any $x\in\mathcal{X}$, if $\|x'-x\|\leq \beta_x$ for $x'\in\mathbb{R}^q$, it holds that
		\begin{equation}
			\|\nabla_x f(x')-\nabla_xf(x)\|\leq \phi_x\|x'-x\|,
		\end{equation}
		where $\phi_x$ and $\beta_x$ are both continuous in $x$.
\item \label{assump-C}%
The function $f$ is coercive in $\mathcal{X}$, i.e., $f(x)\rightarrow+\infty$ if $x\in\mathcal{X}$ satisfies either $\|x\|\rightarrow+\infty$ or $x\rightarrow\partial\mathcal{X}$.
\end{enumerate}
\end{assumption}
	
	The stochastic zeroth-order algorithm aims to solve the stochastic optimization (\ref{optimization}) in the bandit setting, where one only has access to a noisy observation, i.e., the value of $h(x,\xi)$, while the detailed form of $h(x,\xi)$ is unknown. Since the gradient of $f(x)$, i.e., $\nabla_{x}f(x)$, can no longer be computed directly, it will be estimated based on the function value. 
	
	In the literature, the gradient can be estimated by the noisy observations with one-point feedback \cite{Flaxman04,Fazel18,malik2020derivative} or two-point feedback \cite{Agarwal10,Nesterov17,Chen17}. In what follows, we introduce the one-point estimation approach, based on which we will propose our algorithm. The work in this paper is trivially extendable to the two-point feedback case.
	
	Define the unit ball and the $(d-1)$-dimensional sphere (surface of the unit ball) in $\mathbb{R}^d$ as
	\begin{equation}
		\mathbb{B}_d=\{y\in\mathbb{R}^d:\|y\|\leq1\},
	\end{equation}
	and	
	\begin{equation}
		\mathbb{S}_{d-1}=\{y\in\mathbb{R}^d:\|y\|=1\},
	\end{equation}	
	respectively. Given a sample $v\sim\uni(\mathbb{B}_{q})$, define
	\begin{equation}
		\hat{f}(x)=\mathbb{E}_{v\in\mathbb{B}_{q}}[f(x+r v)],
	\end{equation}
	where $r>0$ is called the smoothing radius.
	It is shown in \cite{Flaxman04} that
	\begin{equation}\label{central g}
		\nabla_{x}\hat{f}(x)=\mathbb{E}_{u\in\mathbb{S}_{q-1}}[f(x+r u)u]q/r,
	\end{equation}
	which implies that $f(x+ru)uq/r$ is an unbiased estimate for $\nabla_{x}\hat{f}(x)$. Based on this estimation, a first-order stationary point of (\ref{optimization}) can be obtained by using gradient descent.
	
	\vspace{-0.4cm}
	\subsection{Multi-Agent Stochastic Optimization}
	
	In this paper, we consider the scenario where optimization (\ref{optimization}) is formulated based on a MAS with $N$ agents $\mathcal{V}=\{1,...,N\}$. Let $x=(x_1^{\top},...,x_N^{\top})^{\top}$, where $x_i\in\mathbb{R}^{q_i}$ is the state of agent $i$, and $\sum_{i=1}^Nq_i=q$. Let $\xi=(\xi_1^{\top},...,\xi_N^{\top})^{\top}$, where $\xi_i\in\mathbb{R}^{p_i}$ denotes the noisy exploratory input applied to agent $i$, and $\sum_{i=1}^Np_i=p$. Different agents may have different dimensional states and noise vectors, and each agent $i$ only has access to a local observation $h_i(x_{\mathcal{N}_i},\xi_{\mathcal{N}_i})$, $i=1,...,N$. Here $x_{\mathcal{N}_i}=\{x_j,j\in\mathcal{N}_i\}$ and $\xi_{\mathcal{N}_i}=\{\xi_j,j\in\mathcal{N}_i\}$ are the vectors composed of the state and noise information of the agents in $\mathcal{N}_i=\{j\in\mathcal{V}:(j,i)\in\mathcal{E}\}$, respectively. Note that each vertex in graph \footnote{Here graph $\mathcal{G}$ can be either undirected or directed. In Section \ref{sec MAS} where our algorithm is applied to the LQR problem, graph $\mathcal{G}$ corresponds to the learning graph.} $\mathcal{G}=(\mathcal{V},\mathcal{E})$ is considered to have a self-loop, i.e., $i\in\mathcal{N}_i$. At this moment we do not impose other conditions on $\mathcal{G}$, instead, we give Assumption \ref{as local cost} on those local observations, based on which the distributed RL algorithm will be proposed. In Section \ref{sec MAS}, when applying our RL algorithm to a model-free multi-agent LQR problem, the details about how to design the inter-agent interaction graph for validity of Assumption \ref{as local cost} will be introduced.
	
	\begin{assumption}\label{as local cost}\footnote{A direct consequence of Assumption \ref{as local cost} is that $f_i(x)$ is locally Lipschitz continuous and has a locally Lipschitz continuous gradient w.r.t. $x_i$. Moreover, the two Lipschitz constants are the same as that of $f(x)$.}
		There exist local cost functions $f_i(x)=\mathbb{E}_{\xi\sim\mathcal{D}}[h_i(x_{\mathcal{N}_i},\xi_{\mathcal{N}_i})]$ and a constant $c>0$ such that for any $i\in\mathcal{V}$ and $x\in\mathcal{X}$, $h_i(\cdot,\cdot): \mathbb{R}^{\sum_{j\in\mathcal{N}_i}q_j}\times \mathbb{R}^{\sum_{j\in\mathcal{N}_i}p_j}\rightarrow\mathbb{R}_{\geq0}$ satisfies  
		\begin{equation}\label{hi<cfi}
		    h_i(x_{\mathcal{N}_i},\xi_{\mathcal{N}_i})\leq cf_i(x), \quad \text{almost surely~(a.s.)},
		\end{equation}
		and 
		\begin{equation}\label{grhi}
			\nabla_{x_i}f(x)=\nabla_{x_i}f_i(x).
		\end{equation}
	\end{assumption}
	\begin{remark}	
		Inequality (\ref{hi<cfi}) is used to build a relationship between $h_i(x,\xi)$ and its expectation w.r.t. $\xi$. Note that (\ref{hi<cfi}) is the only condition for the random variable $\xi$, and $\xi$ does not necessarily have a zero mean. If $h_i(x,\xi)$ is continuous in $\xi$ for any $x\in\mathcal{X}$, the assumption (\ref{hi<cfi}) holds when the random variable $\xi$ is bounded. When $\xi$ is unbounded, for example, $\xi$ follows a sub-Gaussian distribution, (\ref{hi<cfi}) may hold with a high probability, see \cite{Rudelson13}. A truncation approach can be used when evaluating the local costs to ensure the boundedness of the observation if $\xi$ follows a standard Gaussian distribution.
	\end{remark}
	
	To show that Assumption \ref{as local cost} is reasonable, we give a class of feasible examples. Consider
	\begin{equation}\label{eq:example}
		h(x,\xi)=\sum_{(i,j)\in\mathcal{E}}F_{ij}(x_i,x_j,\xi_i,\xi_j), ~~~~\xi_i\sim\mathcal{D}_i,
	\end{equation}
	where $F_{ij}(\cdot,\cdot,\cdot,\cdot): \mathbb{R}^{q_i}\times\mathbb{R}^{q_j}\times\mathbb{R}^{p_i}\times\mathbb{R}^{p_j}$ is locally Lipschitz continuous and has a locally Lipschitz continuous gradient w.r.t. $x_i$ and $x_j$, $\mathcal{E}$ is the edge set of a graph characterizing the inter-agent coupling relationship in the objective function, and $\mathcal{D}_i$ for each  $i\in\mathcal{V}$ is a bounded distribution with zero mean. By setting $h_i(x_{\mathcal{N}_i},\xi_{\mathcal{N}_i})=\sum_{j\in\mathcal{N}_i}F_{ij}(x_i,x_j,\xi_i,\xi_j)$, Assumption \ref{as local cost} is satisfied. If we define the domain $\mathcal{X}=\mathcal{X}_1\times\cdots\times\mathcal{X}_N$ as a strategy set, then the formulation \eqref{optimization} with \eqref{eq:example} can be viewed as a cooperative networking game \cite{chkhartishvili2018social}, where $x_i$ is the decision variable of player $i$, each pair of players aim to minimize $F_{ij}$, and $\xi_i$ is the random effect on the observation of each player. Moreover, the objective function \eqref{eq:example} has also been employed in many multi-agent coordination problems \cite{Chen14} such as consensus \cite{Qin16} and formation control \cite{Oh15}. As a specific example, consider $F_{ij}(x_i,x_j,\xi_i,\xi_j) = \|x_i+\xi_i-x_j-\xi_j\|^2$. In this case, the objective is consensus where $\xi_i$ and $\xi_j$ model bounded noise effects in the measurement of $x_i$ and $x_j$, respectively.

	In this paper, with Assumption \ref{as local cost}, we will propose a novel distributed zeroth-order algorithm, where the local gradient of each agent $i$ is estimated based upon the local observation $h_i$. The problem to be solved is summarized as follows.
	\begin{problem}
		Under Assumptions \ref{assumption1}-\ref{as local cost}, given an initial state $x^0\in\mathcal{X}$, design a distributed algorithm for each agent $i$ based on the local observation\footnote{Since $x_{\mathcal{N}_i}$ is composed of partial elements of $x$, for symbol simplicity, we write the local cost function for each agent $i$ as $h_i(x,\xi)$ and treat $h_i(\cdot,\cdot)$ as a mapping from $\mathbb{R}^q\times\mathbb{R}^p$ to $\mathbb{R}_{\geq0}$ while keeping in mind that $h_i(x,\xi)$ only involves agent $i$ and its neighbors.} $h_i(x,\xi)$ such that by implementing the algorithm, the state $x$ converges to a stationary point of $f$.
	\end{problem}

	\section{Distributed ZOO with Asynchronous Samples and Updates}	\label{sec: as}
	In this  section, we propose a distributed zeroth-order algorithm with asynchronous samples and updates based on an accelerated zeroth-order BCD algorithm. 

 \vspace{-0.4cm}
	\subsection{Block Coordinate Descent}
	
	A BCD algorithm solves  for the minimizer $x=(x_1^{\top},...,x_N^{\top})^{\top}\in\mathbb{R}^q$ by updating only one block including partial components of $x$ in each iteration. More specifically, let $\mathcal{I}^B_k\subset\mathcal{V}$ be the index set of the block to be updated at step $k$, and $x^k$ be the value of $x$ at step $k$. An accelerated BCD algorithm is shown below:
	\begin{equation}\label{BCD}
		\left\{
		\begin{array}{lrlr}
			x_i^{k+1}=x_i^{k}, & i\notin \mathcal{I}^B_k,\\
			x_i^{k+1}=\hat{x}_i^k-\eta\nabla_{x_i} f(\hat{x}_i^k,x_{-i}^k), & i\in\mathcal{I}^B_k,
		\end{array}
		\right.
	\end{equation}
	where $\eta>0$ is the step-size,  $\hat{x}_i^k$ is the extrapolation determined by
	\begin{equation}\label{xhati}
		\hat{x}_i^k=x_i^k+w_i^k(x_i^k-x_i^{k_{prev}}),
	\end{equation}	
	here $w_i^k\geq0$ is the extrapolation weight to be determined, $k_{prev}$ is the latest step at which $x_i$ was updated, i.e., $k_{prev}=\max\{k':i\in\mathcal{I}_{k'}^B, k'\leq k\}$.
	
	Note that $w_i^k$ can be simply set as zero. However, it has been empirically shown in \cite{wright2015coordinate,xu2017globally} that having appropriate positive extrapolation weights helps significantly accelerate the convergence speed of the BCD algorithm, which is also observed in our simulation results.
	
	To avoid using the first-order information, which is usually absent in reality, we estimate $\nabla_{x_i} f(\hat{x}_i^k,x_{-i}^k)$ in (\ref{BCD}) based on the observation $f_i(\hat{x}_i^k,x_{-i}^k)$, see the next subsection.
	

	\vspace{-0.4cm}
     \subsection{Gradient Estimation via Local Cost Evaluation}
	 In this subsection, we introduce given $x\in\mathcal{X}$, how to estimate $\nabla_{x_i} f(x)$ based on $f_i(x)$. In the ZOO literature, $\nabla_{x}f(x)$ is estimated by perturbing state $x$ with a vector randomly sampled from $\mathbb{S}_{q-1}$. In order to achieve distributed learning, we expect different agents to sample their own perturbation vectors independently. Based on Assumption \ref{as local cost}, we have
	\begin{equation}\label{fi}
		\nabla_{x}f(x)=\begin{pmatrix}
			\nabla_{x_1}f(x)\\
			\colon\\
			\nabla_{x_N}f(x)
		\end{pmatrix}=\begin{pmatrix}
			\nabla_{x_1}f_1(x)\\
			\colon\\
			\nabla_{x_N}f_N(x)
		\end{pmatrix}.
	\end{equation}

	Let 
	\begin{equation}
		\begin{split}
			\hat{f}_i(x)&=\mathbb{E}_{v_i\in\mathbb{B}_{q_i}}[f_i(x_i+r_i v_i,x_{-i})]\\ &=\frac{\int_{r_i\mathbb{B}_{q_i}}f_i(x_i+ v_i,x_{-i})dv_i}{V(r_i\mathbb{B}_{q_i})},
		\end{split}
	\end{equation}
	where $V(r_i\mathbb{B}_{q_i})$ is the volume of  $r_i\mathbb{B}_{q_i}$. Here $\hat{f}_i(x)$ is always differentiable even when $f_i(x)$ is not differentiable.
	To approximate $\nabla_{x_i}f_i(x)$ for each agent $i$, we approximate $\nabla_{x_i}\hat{f}_i(x)$ by the following one-point feedback:
	\begin{equation}\label{gixuxi}
		g_i(x,u_i,\xi)=\frac{q_i}{r_i}h_i(x_i+r_iu_i,x_{-i},\xi)u_i,
	\end{equation}
 $u_i\in\uni(\mathbb{S}_{q_i-1})$, $\xi\in\mathbb{R}^p$ is a random variable following the distribution $\mathcal{D}$. Note that according to the definition of the local cost function $h_i(x,\xi)$, $g_i(x,u_i,\xi)$ may be only affected by partial components of $\xi$.
	
	The following lemma shows that $g_i(x,u_i,\xi)$ is an unbiased estimate of $\nabla_{x_i}\hat{f}_i(x)$.
	
	\begin{lemma}\label{le fhat=Eg}
		Given $r_i>0$, $i=1,...,N$, the following holds
		\begin{equation}
			\nabla_{x_i}\hat{f}_i(x)=\mathbb{E}_{u_i\in\mathbb{S}_{q_i-1}}\mathbb{E}_{\xi\sim\mathcal{D}}[g_i(x,u_i,\xi)].
		\end{equation}	
	\end{lemma}
	Although $\nabla_{x_i}\hat{f}_i(x)\neq\nabla_{x_i}f(x)$, their error can be quantified using the smoothness property of $f(x)$, as shown below. 	
	\begin{lemma}\label{le fhat error}
		Given a point $x\in\mathbb{R}^{q}$, if $r_i\leq\beta_x$, then
		\begin{equation}\label{g=gradf}
			\|\nabla_{x_i}\hat{f}_i(x)-\nabla_{x_i}f(x)\|\leq\phi_xr_i.
		\end{equation}
	\end{lemma}
	
	\begin{remark}
		Compared with gradient estimation based on one-point feedback, the two-point feedback $g_i(x,u_i,\xi_i)=\frac{q_i}{r_i}\left[h_i(x_i+r_iu_i,x_{-i},\xi_i)-h_i(x_i-r_iu_i,x_{-i},\xi_i)\right]u_i$ is recognized as a more robust algorithm with a smaller variance and a faster convergence rate \cite{Agarwal10,Nesterov17,Chen17}. In this work, we mainly focus on how to solve an optimization via networks by a distributed zeroth-order BCD algorithm. Since the expectation of the one-point feedback is equivalent to the expectation of the two-point feedback, our algorithm in this paper is extendable to the two-point feedback case. Note that in the gradient estimation, the two-point feedback requires two times of policy evaluation with the same noise vector, which may be unrealistic in practical applications.
	\end{remark}	
	
	\vspace{-0.4cm}
	\subsection{Distributed ZOO Algorithm with Asynchronous Samplings}
	In this subsection, we propose a distributed ZOO algorithm with asynchronous sample and update schemes based on the BCD algorithm (\ref{BCD}) and the gradient approximation for each agent $i$. According to (\ref{gixuxi}), we  have the following approximation for each agent $i$ at step $k$:
	\begin{equation}\label{gixhat}
		g_i(\hat{x}_i^k, x_{-i}^k, u_i^k,\xi)=\frac{q_i}{r_{i}}h_i(\hat{x}_i^k+r_{ik}u_i^k,x_{-i}^k,\xi^k)u_i^k,
	\end{equation}
	where $u_i^k$ is uniformly randomly sampled from $\mathbb{S}_{q_i-1}$.
	
	In fact, since $h_i(x,\xi)$ only involves the agents in $\mathcal{N}_i$, it suffices to maintain the variables of the agents in $\mathcal{N}_i\setminus\{i\}$ invariant when estimating the gradient of agent $i$. That is, in our problem, two agents are allowed to update their variables simultaneously if they are not neighbors. This is different from a standard BCD algorithm where only one block of the entire variable is updated in one iteration.
	
	To achieve simultaneous update for non-adjacent agents, we decompose the set of agents $\mathcal{V}$ into $s$ independent clusters ($s$ has an upper bound depending on the graph), i.e.,  $\mathcal{V}=\cup_{j=1}^s\mathcal{V}_j$, and 
	\begin{equation}
\mathcal{V}_{j_1}\cap\mathcal{V}_{j_2}=\varnothing, ~~\forall \text{ distinct } j_1,j_2\in\{1,...,s\}
	\end{equation}
	such that the agents in the same cluster are not adjacent in the interaction graph, i.e., $(i_1,i_2)\notin\mathcal{E}$ for any $i_1,i_2\in\mathcal{V}_{j}$, $j\in\{1,...,s\}$. Note that no matter graph $\mathcal{G}$ is directed or undirected, any two agents  have to lie in different clusters if there is a link from one to the other. Without loss of generality, we assume $\mathcal{G}$ is undirected, and let $\mathcal{N}_i$ be the neighbor set of agent $i$. In the case when $\mathcal{G}$ is directed, we define $\mathcal{N}_i$ as the set of agents $j$ such that $(i,j)\in\mathcal{E}$ or $(j,i)\in\mathcal{E}$. A simple algorithm for achieving such a clustering is shown in Section \ref{app alg}. Based on the clustering, we propose Algorithm \ref{alg:as} as the asynchronous distributed zeroth-order  algorithm. Algorithm \ref{alg:as} can be viewed as a distributed RL algorithm where different clusters take actions asynchronously, different agents in one cluster take actions simultaneously and independently. In Algorithm \ref{alg:as}, step 5 can be viewed as {\it policy evaluation} for agent $i$, while step 6 corresponds to {\it policy iteration}. Moreover, the local observation $h_{i}(\hat{x}_i^k+r_iu_{i}^k,x_{-i}^k,\xi^k)$ can be viewed as the reward returned by the environment to agent $i$.
	\begin{algorithm}[htbp]
		\small
		\caption{Distributed Zeroth-Order Algorithm with Asynchronous Samplings}\label{alg:as}
		\textbf{Input}: Step-size $\eta$, smoothing radius $r_i$ and variable dimension $q_i$,  $i=1,...,N$, clusters $\mathcal{V}_j$, $j=1,...,s$, iteration number $T$, update order $z_k$ (the index of the cluster to be updated at step $k$) and extrapolation weight $w_i^k$, $k=0,...,T-1$,  initial point $x^0\in\mathcal{X}$.\\
		\textbf{Output}: $x(T)$.
		\begin{itemize}
			\item[1.] \textbf{for} $k=0,1,...,T-1$ \textbf{do}
			\item[2.] ~~~~Sample $\xi^k\sim\mathcal{D}$.
			\item[3.] ~~~~\textbf{for} all $i\in\mathcal{V}$ \textbf{do} 
			\item[4.]~~~~~~~~\textbf{if} $i\in\mathcal{V}_{z_k}$ \textbf{do} (Simultaneous Implementation)
			\item[5.] ~~~~~~~~~~~~Agent $i$ computes $\hat{x}_i^k$ by (\ref{xhati}), samples $u_{i}^k$ randomly from $\mathbb{S}_{q_i-1}$ and observes  $h_{i}(\hat{x}_i^k+r_iu_{i}^k,x_{-i}^k,\xi^k)$.
			\item[6.]~~~~~~~~~~~~Agent $i$ computes the estimated local gradient $g_i(\hat{x}_i^k,x_{-i}^k,u_i^k,\xi^k)$ according to (\ref{gixhat}). Then updates its policy:
			\begin{equation}
				x_i^{k+1}=\hat{x}_i^k-\eta g_i(\hat{x}_i^k,x_{-i}^k,u_i^k,\xi^k).
			\end{equation}
		   \item[7.]~~~~~~~~\textbf{else}
		  \item[8.]\begin{equation}
			x_i^{k+1}=x_i^k. 
		  \end{equation}
		
		    \item[9.]~~~~~~~~\textbf{end}
		    \item[10.] ~~~~\textbf{end}
			\item[11.] \textbf{end}
		\end{itemize}
	\end{algorithm}	
	
	In the literature, BCD algorithms have been studied with different update orders such as deterministically cyclic \cite{canutescu2003cyclic,wright2015coordinate} and randomly shuffled \cite{richtarik2014iteration,yu2019zeroth}. In this paper, we adopt an ``essentially cluster cyclic update" scheme, which includes the standard cyclic update as a special case, and is a variant of the essentially cyclic update scheme in \cite{wright2015coordinate,xu2017globally}.
	
	\begin{assumption}\label{as cyclic}(Essentially Cluster Cyclic Update)
		Given integer $T_0\geq s$, for any cluster $j\in\{1,...,s\}$ and any two steps $k_1$ and $k_2$ such that $k_2-k_1=T_0-1$, there exists $k_0\in [k_1,k_2]$ such that $z_{k_0}=j$.
	\end{assumption}
	
	Assumption \ref{as cyclic} implies that each cluster of agents update their states at least once during every consecutive $T_0$ steps. When $|\mathcal{V}_i|=1$ for $i=1,...,s$, and $s=N$, Assumption \ref{as cyclic} implies an essentially cyclic update in \cite{wright2015coordinate,xu2017globally}.

	To better understand Algorithm \ref{alg:as}, let us look at a multi-agent coordination example.
	
	\begin{example}\label{ex1}
		Suppose that the global function to be minimized is (\ref{eq:example}) with  $\mathcal{E}=\{(1,2),(2,3),(3,4)\}$ being the edge set. Then the local cost function for each agent is:
		\begin{equation}
			\begin{split}
				&h_1=F(x_1,x_2,\xi_{12}),~~ h_2=F(x_1,x_2,\xi_{12})+F(x_2,x_3,\xi_{23}),\\
				&h_3=F(x_2,x_3,\xi_{23})+F(x_3,x_4,\xi_{34}), ~~h_4=F(x_3,x_4,\xi_{34}),
			\end{split}
		\end{equation}
		where $\xi_{ij}$ is a bounded zero-mean noise. By implementing Algorithm \ref{alg:clustering}, the two clusters obtained are $\mathcal{V}_1=\{1,3\}$ and $\mathcal{V}_2=\{2,4\}$. By implementing Algorithm \ref{alg:as} with two clusters updating alternatively, the diagram for two consecutive iterations is shown in Fig. \ref{fig diagram as}, where agents in $\mathcal{V}_1$ and $\mathcal{V}_2$ take their actions successively.


		\begin{figure}
			\centering
			\includegraphics[width=9cm]{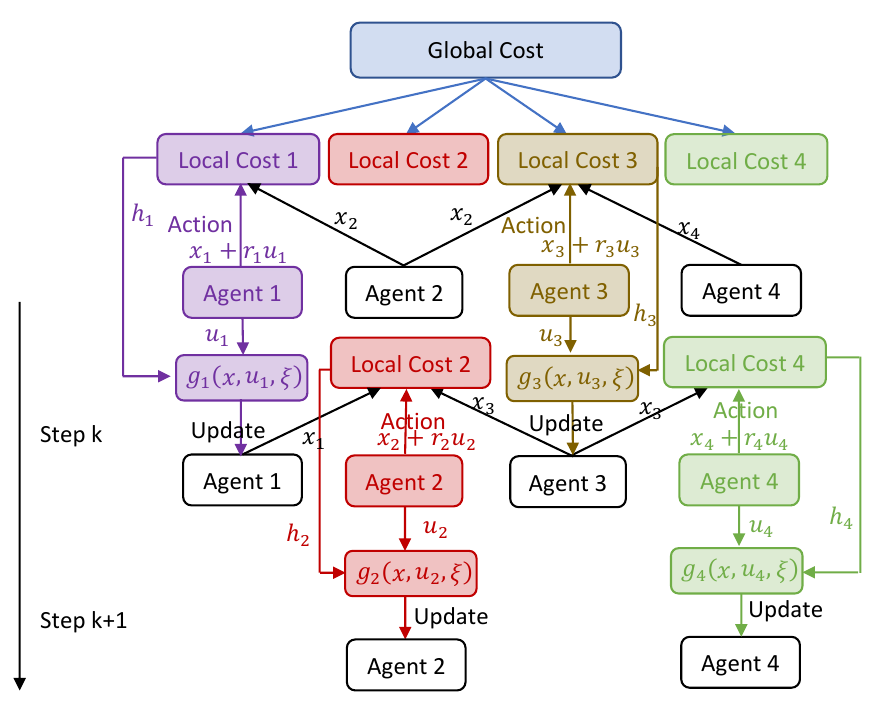}
			\caption{The architecture of distributed RL via asynchronous actions during two consecutive iterations.} \label{fig diagram as}
		\end{figure}
	\end{example}
	
	\vspace{-0.5cm}
	\subsection{Convergence Result}
	
	We study the convergence of Algorithm \ref{alg:as} by focusing on $x\in\mathbb{X}$, where $\mathbb{X}$ is defined as 
	\begin{multline}\label{Xx}
		\mathbb{X}=\{x\in\mathbb{R}^{nN}:f(x)\leq\alpha f(x^0),\\ f_i(x)\leq\alpha_if_i(x^0), \forall~ i\in\mathcal{V}\}\subseteq \mathcal{X},
	\end{multline}
	where $\alpha, \alpha_i>1$, $i\in\mathcal{V}$, $x^0\in\mathbb{X}$ is the given initial condition. 
	Since $f(x)$ is continuous and coercive, the set $\mathbb{X}$ is compact. Then we are able to find uniform parameters feasible for $f(x)$ over $\mathbb{X}$:
    \begin{equation}
    	\phi_0=\sup_{x\in\mathbb{X}}\phi_x,~~\lambda_0=\sup_{x\in\mathbb{X}}\lambda_x, ~~\rho_0=\inf_{x\in\mathbb{X}}\{\beta_x,\zeta_x\}.
    \end{equation}	
	The following theorem shows an approximate convergence result based on establishing the probability of the event $\{x^k\in\mathbb{X}\}$ for $k=0,...,T-1$. Let $N_i$ denote the number of agents in the cluster containing agent $i$. For notation simplicity, we denote $N_0=\max_{i\in\mathcal{V}}N_i=\max_{j\in\{1,...,s\}}|\mathcal{V}_j|$, $q_+=\max_{i\in\mathcal{V}}q_i$, $r_-=\min_{i\in\mathcal{V}}r_i$, $f_0(x^0)=\max_{i\in\mathcal{V}}\alpha_if_i(x^0)$. 
	\begin{theorem}\label{th as}
		Under Assumptions \ref{assumption1}-\ref{as local cost}, given positive scalars $\epsilon$, 
		$\nu$, $\gamma$, and $\alpha\geq 2+\gamma+\frac{1}{\nu}+\nu\gamma$, $x^0\in\mathcal{X}$. Let $\{x^k\}_{k=0}^{T-1}$ be a sequence of states obtained by implementing Algorithm \ref{alg:as} for $k=0,...,T-1$. Suppose that\footnote{Note that the condition on $w_i^k$ in Theorem 1 actually imposes an upper bound for $\|w_i^k(x_i^k-x_i^{k_{prev}})\|$. This condition is naturally satisfied when $x_i^k-x_i^{k_{prev}}=0$.}

 \begin{equation}\label{conditions}
		\begin{split}
		    &T=\lceil \frac{2\alpha \nu f(x^0)}{\eta\epsilon}\rceil, ~~~r_i\leq \min\{\frac{\rho_0}{2},\frac{1}{2\phi_0}\sqrt{\frac{\gamma\epsilon}{\alpha N_0}}\},\\
		    &\eta\leq\min\{\frac{\rho_0}{2\delta\sqrt{N_0}},\frac{2\alpha f(x^0)}{\gamma\epsilon},\frac{\gamma\epsilon}{2\alpha N_0(\phi_0\delta^2+4\phi_0^2+\phi_0+4)}\},\\
		    &w_i^k\leq\frac{1}{\|x_i^k-x_i^{k_{prev}}\|}\min\{\eta^{3/2},\frac{\rho_0}{2\sqrt{N_i}}\},~~~  i\in\mathcal{V},
		\end{split}
		\end{equation}
where $\delta=\frac{q_+}{r_-}c\left[f_0(x^0)+\lambda_0\rho_0\right]$ is the uniform bound on the estimated gradient (as shown in Lemma \ref{le E[g]}). The following statements hold.
		
		(i). The following inequality holds with a probability at least $1-\frac{1}{\alpha}(2+\gamma+\frac{1}{\nu}+\nu\gamma)$:
		\begin{equation}
			\frac{1}{T}\sum_{k=0}^{T-1}\sum_{i\in\mathcal{V}_{z_k}}\|\nabla_{x_i}f(x^k)\|^2< \epsilon.
		\end{equation}	
	(ii). Under Assumption \ref{as cyclic}, suppose that there exists some $\bar{\epsilon}\in(0,\rho_0]$ such that
	\begin{equation}\label{extra conditions}
	w_i^k\leq \frac{\bar{\epsilon}}{2(T_0-1)N_i\|x_i^k-x_i^{k_{prev}}\|},~~ \eta\leq\frac{\bar{\epsilon}}{2\delta (T_0-1)N_0},
	\end{equation}
	for any step $k\in\{0,..., T-1\}$, and conditions in (\ref{conditions}) are satisfied,  then the following holds with a probability at least $1-\frac{1}{\alpha}(2+\gamma+\frac{1}{\nu}+\nu\gamma)$:
 \begin{equation}
	    \frac{1}{T}\sum_{k=0}^{T-1}\|\nabla_xf(x^k)\|^2<\hat{\epsilon} .
	\end{equation}
	where $\hat{\epsilon}=2T_0(\epsilon+\phi_0^2\bar{\epsilon}^2)$.
	\end{theorem}
	
	\textbf{Approximate Convergence.} Given positive scalars $\epsilon$ $\alpha$, $\gamma$, and $\nu$, Theorem \ref{th as} (i) implies that Algorithm \ref{alg:as} converges to a sequence $\{x_k\}$, where the partial derivative of $x_k$ w.r.t. one cluster is close to 0 at each step with high probability. When $s=1$, Algorithm \ref{alg:as} reduces to an accelerated zeroth-order gradient descent algorithm, and Theorem (i) implies convergence to a $\epsilon$-accurate stationary point with high probability. Theorem \ref{th as} (ii) implies that under an ``essentially cluster cyclic update" scheme, Algorithm \ref{alg:as} converges to a $\hat{\epsilon}$-accurate stationary point if the extrapolation weight $w_i^k$ and the step-size $\eta$ are both sufficiently small.
	
	\textbf{Convergence Probability.}	The convergence probability can be controlled by adjusting parameters $\alpha$, $\gamma$ and $\nu$. Similar claims have been made in the literature (see e.g.,~\cite[Theorem 1]{Li19}). However, instead of providing an explicit probability, we can use $\nu$ and $\gamma$ to adjust the probability. For example, set $\alpha=20$, $\gamma=1$, $\nu=1$, then the probability is at least $1-1/4=3/4$. To achieve a high probability of convergence, it is desirable to have large $\alpha$ and $\nu$, and small $\gamma$, which implies that the performance of the algorithm can be enhanced at the price of adopting a large number of samples. As shown in~\eqref{conditions}, for a given $\epsilon$ and a sufficiently large $\alpha$, $\gamma$ is used to control the step-size $\eta$ and the sampling radius $r_i$ while $\nu$ is used to control the total number of iterations.

	\textbf{Sample Complexity.} To achieve high accuracy of the convergent result, we consider $\epsilon$ as a small positive scalar such that $\epsilon\ll1/\epsilon$. As a result, the sample complexity for convergence of Algorithm \ref{alg:as} is $T\sim\mathcal{O}(q_+^2N_0^2/\hat{\epsilon}^3)$. This implies that the required iteration number mainly depends on the highest dimension of the variable for one agent, the largest size of one cluster, and the number of clusters (because $T_0\geq s$). Note that even when the number of agents increases, $q_+$ may remain the same, implying high scalability of our algorithm to large-scale networks\footnote{The scalability advantage of our algorithm compared with a relevant model-free distributed LQR reference will be given in the analysis after Theorem \ref{th as LQR}.}. Moreover, $N_0$ may increase as the number of clusters $s$ decreases. Hence, there is a trade-off between the benefits of minimizing the largest cluster size against minimizing the number of clusters. When the network is of large scale, $T_0$ dominates the sample complexity, which makes minimizing the number of clusters the optimal clustering strategy. Note that the sample complexity is directly associated with the convergence accuracy.
	Analysis with diminishing step-sizes is of interest and may help achieve asymptotic convergence of the gradient.

	Theorem~\ref{th as} provides sufficient conditions for Algorithm~\ref{alg:as} to converge. The objective of our analysis is to identify the range of applicability of the proposed algorithm, and establish the qualitative behavior of the algorithms, rather than providing optimal choices of the parameters for the algorithm. Some of the parameters, such as the Lipschitz constants, have analytical expressions in the model-based setting (see e.g.,~\cite{Bu19}). However, they are  difficult to obtain in the model-free setting. In our experiments, we choose small step-sizes empirically to ensure that the cost function decreases over time.
  \vspace{-0.4cm}
	\subsection{Variance Analysis}\label{subsec: as variance}
	
	In this subsection, we analyze the variance of our gradient estimation strategy and make comparisons with the gradient estimation via global cost evaluation. Without loss of generality, we analyze the estimation variance for the $i$-th agent. Let $u_i\sim\text{Uni}(\mathbb{S}_{q_i-1})$, $z\sim\text{Uni}(\mathbb{S}_{q-1})$, and $z_i\in\mathbb{R}^{q_i}$ be a component of $z$ corresponding to agent $i$. Under the same smoothing radius $r>0$, one time gradient estimates based on the local cost $h_i$ and the global cost $h$ are
	\begin{equation}\label{g_las}
		g_l=\frac{q_i}{r}h_i(x_i+ru_i,x_{-i},\xi)u_i,
	\end{equation}
	and
	\begin{equation}\label{g_g}
		g_{g}=\frac{q}{r}h(x+rz,\xi)z_i,
	\end{equation}
	respectively.
	
	\begin{lemma}\label{le cov g as g}
		The covariance matrices for the gradient estimates (\ref{g_las}) and (\ref{g_g}) are
		\begin{multline}\label{eq:cov1}
			\Cov(g_l) = \frac{q_i}{r^2}\Big[\mathbb{E}[h_i^2(x, \xi)]I_{q_i} \\
			- \mathbb{E}[\nabla_{x_i} h_i(x,\xi)]\mathbb{E}^{\top} [\nabla_{x_i}h_i(x,\xi)]+\mathcal{O}(r^2)\Big], 
		\end{multline}
		and	\begin{multline}
			\Cov(g_g) = \frac{q}{r^2}\Big[\mathbb{E}[h^2(x,\xi)]I_{q_i} \\
			-\mathbb{E}[\nabla_{x_i}h(x,\xi)]\mathbb{E}^{\top}[\nabla_{x_i}h(x,\xi)]+\mathcal{O}(r^2)\Big], \label{eq:cov2}
		\end{multline}
		respectively.
	\end{lemma}

	Due to Assumption \ref{as local cost}, the difference between two covariance matrices is
	\begin{multline}\label{variance difference}
		\Cov(g_g)-\Cov(g_l)=\frac{1}{r^2}\Big[\sum_{j\in\mathcal{V}\setminus\{i\} }q_j\mathbb{E}[h^2(x,\xi)]I_{q_i}\\+q_i\mathbb{E}[h^2(x,\xi)-h_i^2(x,\xi)]I_{q_i}+\mathcal{O}(r^2)\Big],
	\end{multline}
	where the first term is positive definite as long as there are more than one agent in the network, the second term is usually positive semi-definite, and the third term is negligible if $r$ is much smaller than $h(x,\xi)$.
	
	Observe that the first two terms may have extremely large traces if the network is of large scale. This implies that the gradient estimate (\ref{g_las}) leads to a high scalability of our algorithm to large-scale network systems.

	\section{Application to Distributed RL of Model-Free Distributed Multi-Agent LQR}\label{sec MAS}
	
	In this section, we will show how Algorithm \ref{alg:as} can be applied to distributively learning a sub-optimal distributed controller for a linear MAS with unknown dynamics.

 \vspace{-0.4cm}
	\subsection{Multi-Agent LQR}
	
	Consider the following MAS with decoupled agent dynamics\footnote{Decoupled agent dynamics are commonly assumed in MASs; for example, a group of robots. Our results are extendable to the case of coupled dynamics. In that case, the coupling relationship in agents' dynamics has to be taken into account in the design of the local cost function and the learning graph in next subsection. Details are explained in Remark \ref{re coupled dynamics}.}:
	\begin{equation}\label{MAS}
		x_i(t+1)=A_ix_i(t)+B_iu_i(t), ~~~~i=1,..., N
	\end{equation}
	where $x_i\in\mathbb{R}^n$, $u_i\in\mathbb{R}^m$, $A_i\in\mathbb{R}^{n\times n}$ and $B_i\in\mathbb{R}^{n\times m}$ are assumed to be unknown. The entire system dynamics becomes
	\begin{equation}\label{hom MAS}
		x(t+1)=\mathcal{A}x(t)+\mathcal{B}u(t),
	\end{equation}
    where $\mathcal{A}\in\mathbb{R}^{nN\times nN}$ and $\mathcal{B}\in\mathbb{R}^{nN\times mN}$ are block diagonal matrices composed of $\{A_i\}_{i=1}^N$ and $\{B_i\}_{i=1}^N$, respectively, i.e., $\mathcal{A}=\text{blkdiag}\{A_1,...,A_N\}$, $\mathcal{B}=\text{blkdiag}\{B_1,...,B_N\}$.
    
	By considering random agents' initial states, we study the following LQR problem:
	\begin{equation}\label{LQR no noise}
		\begin{split}
			\min_{K}~~~~&J(K)=\mathbb{E}\left[\sum_{t=0}^\infty\gamma^t x^{\top}(t)(Q+K^{\top}RK)x(t)\right]\\
			\text{s.t.}&~~~~x(t+1)=(\mathcal{A}-\mathcal{B}K)x(t), ~~~~x(0)\sim\mathcal{D},
		\end{split}
	\end{equation}
	here $0<\gamma\leq1$, $\mathcal{D}$ is a distribution such that $x(0)$ is bounded and has a positive definite second moment $\Sigma_x=\mathbb{E}[x(0)x^\top(0)]$, $Q=G\odot\tilde{Q}$ with $G\in\mathbb{R}^{N\times N}\succeq0$, $\tilde{Q}\in\mathbb{R}^{nN\times nN}\succeq0$, the symbol $\odot$ denoting the Khatri-Rao product, $R=\diag\{R_1,...,R_N\}\succ0$, and $K\in\mathbb{R}^{mN\times nN}$ is the linear control gain to be designed. Note that while we assume that the agents' dynamics are unknown, the matrices $Q$ and $R$ are considered known. This is reasonable because in reality $Q$ and $R$ are usually artificially designed. It has been shown in \cite{malik2020derivative} that when $0<\gamma<1$, the formulation (\ref{LQR no noise}) is equivalent to the LQR problem with fixed agents' initial states and additive noises in dynamics, where the noise follows the distribution $\mathcal{D}$. Hence, our results are extendable to LQR with noisy dynamics.
	
	When $\gamma\in(0,1)$, let $y(0)=x(0)$, and $y(t+1)=\sqrt{\gamma}(\mathcal{A}-\mathcal{B}K)y(t)$, then we have $y(t)=\gamma^{t/2}x(t)$. It follows that $J(K)=\mathbb{E}[\sum_{t=0}^\infty y^\top(t)(Q+K^\top RK)y(t)]$. This implies that $J(K)$ remains the same as that for $\gamma=1$ by replacing $\mathcal{A}$ and $\mathcal{B}$ in system dynamics (\ref{hom MAS}) with $\sqrt{\gamma}\mathcal{A}$ and $\sqrt{\gamma}\mathcal{B}$.  Hence, we define the following set:
	\begin{equation}
		\mathbb{K}_s=\{K\in\mathbb{R}^{mN\times nN}: \sqrt{\gamma}(\mathcal{A}-\mathcal{B}K)~\text{is Schur Stable}\}.
	\end{equation}
	Note that any $K\in\mathbb{K}_s$ always renders $J(K)$ finite because a Schur stabilizing gain always renders $J(K)$ with $\gamma=1$ finite. Based on \cite[Lemma 3.7]{Bu19} and \cite[Lemmas 4, 5]{malik2020derivative}, we list several properties of $J(K)$ in the following lemma.
	\begin{lemma}\label{lem:a2a3}
		The cost function $J(K)$ in (\ref{LQR no noise}) has the following properties:
		
		(i) $J(K)$ is coercive in the sense that $J(K)\rightarrow\infty$ if $K\in\mathbb{K}_s$ satisfies either $\|K\|\rightarrow\infty$ or $K\rightarrow\partial\mathbb{K}_s$;

		(ii) For any $K\in\mathbb{K}_s$, there exist continuous positive parameters $\lambda_K$, $\zeta_K$, $\phi_K$ and $\beta_K$ such that the cost function $J(K)$ in (\ref{LQR no noise}) is $(\lambda_K,\zeta_K)$ locally Lipschitz continuous and has a $(\phi_K,\beta_K)$ locally Lipschitz continuous gradient.
	\end{lemma}

Based on the matrix $G$, we define the {\it cost graph} interpreting inter-agent coupling relationships in the cost function.
	\begin{definition}(\textbf{Cost Graph})
		The cost graph $\mathcal{G}_C=(\mathcal{V},\mathcal{E}_C)$ is an undirected graph such that $G_{ij}\neq0$ if and only if $(i,j)\in\mathcal{E}_C$. The neighbor set of agent $i$ in the cost graph is defined as $\mathcal{N}_C^i=\{j\in\mathcal{V}: (i,j)\in\mathcal{E}_C\}$. 
	\end{definition}
	
	Distributed control implies that each agent only needs to sense state information of its local neighbors. Next we define {\it sensing graph} interpreting required inter-agent sensing relationships for distributed control.
	
	\begin{definition}(\textbf{Sensing Graph})
		The sensing graph $\mathcal{G}_S=(\mathcal{V},\mathcal{E}_S)$ is a directed graph with each agent having a self-loop. The neighbor set for each agent $i$ in graph $\mathcal{G}_S$ is defined as $\mathcal{N}_S^i=\{j\in\mathcal{V}: (j,i)\in\mathcal{E}_S\}$, where $(j,i)\in\mathcal{E}_S$ implies that agent $i$ has access to $x_j$.
	\end{definition}
	
	\textbf{Notes about the cost graph and the sensing graph}. 
	\begin{itemize}

		\item The cost graph $\mathcal{G}_C$ is determined by the prespecified cost function, and is always undirected becuase $Q$ is positive semi-definite.
		
		\item We assume $\mathcal{G}_C$ is connected. Note that if $\mathcal{G}_C$ is disconnected, then the performance index in (\ref{LQR no noise}) can be naturally decomposed according to those connected components, and the LQR problem can be transformed to smaller sized LQR problems.
		
		\item In real applications, the sensing graph is designed based on the sensing capability of each agent. It is even not necessarily weakly connected.
		
		\item Here the cost graph $\mathcal{G}_C$ and the sensing graph $\mathcal{G}_S$ are defined independently. In specific applications, they can be either related to or independent of each other.
		
	\end{itemize}

	Let $X(i,j)\in\mathbb{R}^{m\times n}$ be a submatrix of $X\in\mathbb{R}^{mN\times nN}$ consisting of elements of $X$ on $(i-1)m+1$-th to $im$-th rows and $(j-1)n+1$-th to $jn$-th columns. The space of distributed controllers is then defined as
	\begin{equation}\label{K space}
		\mathbb{K}_d=\{X\in\mathbb{R}^{mN\times nN}: X(i,j)=\mathbf{0}_{m\times n}~\text{if}~j\notin\mathcal{N}_S^i,~ i,j\in\mathcal{V}\}.
	\end{equation}
	We make the following assumption to guarantee that the distributed LQR problem is solvable.
	\begin{assumption}\label{as solvable}
		$\mathbb{K}_d\cap\mathbb{K}_s\neq\varnothing$.
	\end{assumption}
	
	We aim to design a distributed RL algorithm for agents to learn a sub-optimal distributed controller $K^*\in\mathbb{K}_d$ such that during the learning process, each agent only requires information from partial agents (according to the sensing graph), and takes actions based on the obtained information.
	
	\vspace{-0.4cm}
	\subsection{Local Cost Function and Learning Graph Design}
	
	We have verified Assumption \ref{assumption1} in Lemma \ref{lem:a2a3}.
	To apply Algorithm \ref{alg:as}, it suffices to find local cost functions such that Assumption \ref{as local cost} holds. In this subsection, we propose an approach to design of such local cost functions. 
	
	
	
	Note that the cost function can be written as a function of $K$:
\begin{footnotesize}
 \begin{equation}
		\begin{split}&J(K)=\mathbb{E}\left[\sum_{t=0}^\infty\gamma^t x^{\top}(t)(Q+K^{\top}RK)x(t)\right]\\
			&=\mathbb{E}\left[\sum_{t=0}^\infty\gamma^t x^{\top}(0)(\mathcal{A}-\mathcal{B}K)^{t\top}(Q+K^{\top}RK)(\mathcal{A}-\mathcal{B}K)^tx(0)\right].
		\end{split}
	\end{equation}
 \end{footnotesize}	
	Let $K=[K_1^{\top},...,K_N^{\top}]^{\top}\in\mathbb{R}^{mN\times nN}$. Then $K_i\in\mathbb{R}^{m\times nN}$ is the local gain matrix to be designed for agent $i$. Based on the definition of $\mathbb{K}_d$ in (\ref{K space}), the  distributed controller for each agent $i$ has the form:
	\begin{equation}\label{u_i}
		u_i=-K_ix=-\tilde{K}_ix_{\mathcal{N}_S^i}, 
	\end{equation} 
	where $\tilde{K}_i\in\mathbb{R}^{m\times n_i}$ with $n_i=|\mathcal{N}_S^i|n$.
	We now view the control gain $K_i$ for each agent $i$ as the optimization variable. According to Assumption \ref{as local cost}, we need to find a local cost $J_i(K)$ for each agent $i$ such that its gradient is the same as the gradient of the global cost w.r.t. $K_i$. That is, $\nabla_{K_i}J_i(K)=\nabla_{K_i}J(K)$.
	
	To design the local cost $J_i$ for each agent $i$, we define the following set including all the agents whose control inputs and states will be affected  by agent $i$ during the implementation of the distributed controller:
	\begin{equation}
		\mathcal{V}_S^i=\{j\in\mathcal{V}: \text{A path from}~ i~ \text{to}~ j~ \text{exists in } \mathcal{G}_S\}.
	\end{equation}
	Since different agents are coupled in the cost function, when extracting the local cost function involving an agent $j\in\mathcal{V}_S^i$ from the entire cost function, all of its neighbors in the cost graph (i.e., $\mathcal{N}_C^j$) should be taken into account. Based on the set $\mathcal{V}_S^i$ for each agent $i$,  we formulate the following feasibility problem for each agent $i\in\mathcal{V}$:
	\begin{equation}\label{SDP grad}
		\begin{split}
			&~~~\text{find}~~~ M_i\in\mathbb{R}^{N\times N}\\
			\text{s.t.}~M_i[j,k]&=G_{jk}, \text{ for all } k\in\mathcal{N}_C^j, j\in\mathcal{V}_S^i, \\
			M_i[j,k]&=0,~~ k\in\mathcal{V}\setminus\cup_{j\in\mathcal{V}_S^i}\mathcal{N}_C^j,
		\end{split}
	\end{equation}
	where $M_i[j,k]$ is the element of matrix $M_i$ on the $j$-th row and $k$-th column.
	
	The solution $M_i$ to (\ref{SDP grad}) must satisfy
	\begin{equation}
		\frac{\partial (x^{\top}(M_i\odot \tilde{Q})x)}{\partial x_j}=\frac{\partial (x^{\top}(G\odot\tilde{Q})x)}{\partial x_j} \text{ for all } j\in\mathcal{V}_S^i. 
	\end{equation}
	Moreover, we observe that the solution $M_i\in\mathbb{R}^{N\times N}$ is actually the matrix with the same principal submatrix associated with $\cup_{j\in\mathcal{V}_S^i}\mathcal{N}_C^j$ as $G$, and all the other elements of $M_i$ are zeros. Then $M_i\succeq0$ because all the principal minors of $M_i$ are nonnegative. 
	
	Now, based on the cost graph $\mathcal{G}_C$ and the sensing graph $\mathcal{G}_S$, we give the definition for the communication graph required in distributed learning.
	\begin{definition}(\textbf{Learning Graph})\label{de learning graph}
		The learning-required communication graph $\mathcal{G}_L=(\mathcal{V},\mathcal{E}_L)$ is a directed graph with the edge set $\mathcal{E}_L$ defined as
		\begin{equation}
			\mathcal{E}_L=\{(k,i)\in\mathcal{V}\times\mathcal{V}: k\in\cup_{j\in\mathcal{V}_S^i}\mathcal{N}_C^j, i\in\mathcal{V}\}.\label{eq:learning_graph}
		\end{equation}
		The neighbor set for each agent $i$ in graph $\mathcal{G}_L$ is defined as $\mathcal{N}_L^i=\{k\in\mathcal{V}: (k,i)\in\mathcal{E}_L\}$, where $(k,i)\in\mathcal{E}_L$ implies that there is an oriented link from $k$ to $i$.
	\end{definition} 
	
\begin{figure}
		\centering
		\includegraphics[width=9cm]{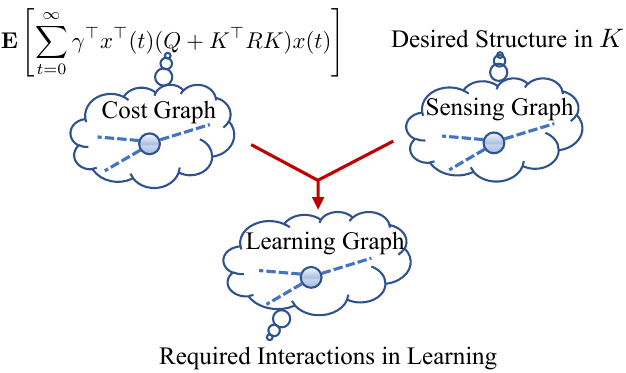}
		\caption{Summary of definitions for the cost, sensing, and learning graphs.} \label{fig graph relationship}
	\end{figure}

\begin{figure}
		\centering
		\includegraphics[width=9cm]{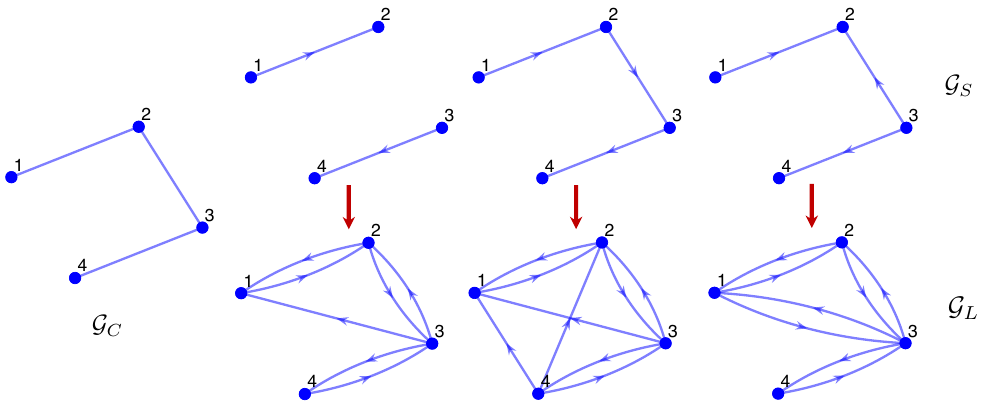}
		\caption{With the same cost graph, three different sensing graphs result in three different learning graphs. Each node in each graph has a self-loop, which is omitted in this figure.} \label{fig graph example}
	\end{figure}

	To better understand the three different graphs, we summarize their definitions in Fig. \ref{fig graph relationship}, and show three examples demonstrating the relationships between $\mathcal{G}_S$, $\mathcal{G}_C$ and $\mathcal{G}_L$ in Fig. \ref{fig graph example}.
	
	\begin{remark}
		There are two points we would like to note for the learning graph. First, since we consider that all the agents have self-loops, by Definition \ref{de learning graph}, each edge of graph $\mathcal{G}_C$ must be an edge of $\mathcal{G}_L$. Second, when $\mathcal{G}_S$ is strongly connected, $\mathcal{G}_L$ is a complete graph because in this case $\mathcal{V}_S^i=\mathcal{V}$ for all $i\in\mathcal{G}$. As a result, if we regard each node in a sensing graph in Fig. \ref{fig graph example} as a strongly connected component, then the resulting learning graph $\mathcal{G}_L$ still has the same structure as it is shown in Fig. \ref{fig graph example}, where each node denotes a fully connected component, and the edge from node $a$ to another node $b$ denotes edges from all agents in component $a$ to all agents in component $b$.
	\end{remark}
	
	\begin{remark}\label{re coupled dynamics}
		When the MAS has coupled dynamics, there will be another graph $\mathcal{G}_{D}=(\mathcal{V},\mathcal{E}_{D})$ describing the inter-agent coupling in dynamics. We assume that $\mathcal{G}_{D}$ is known, which is reasonable in many model-free scenarios because we only need to have the coupling relationship between different agents. In this scenario, to construct the learning graph, the sensing graph used in defining (\ref{eq:learning_graph}) should be replaced by a new graph $\mathcal{G}_{SD}=(\mathcal{V},\mathcal{E}_{SD})$ with $\mathcal{E}_{SD}=\mathcal{E}_{D}\cup\mathcal{E}_S$, where $\mathcal{E}_S$ is the edge set of the graph $\mathcal{G}_S$ describing the desired structure of the distributed control gain. If $\mathcal{E}_{D}\subseteq\mathcal{E}_S$, then the learning graph is the same for the multi-agent networks with coupled and decoupled dynamics. The learning graph describes the required information flow between the agents. However, the learning process may be distributed by consensus-based estimation to avoid having a dense communication graph. Another way to reduce communication is to redesign the local cost of each agent by ignoring agents that are far away from it in $\mathcal{G}_D$. In \cite{qu2020scalable}, it has been shown that ignoring those agents beyond the $\kappa$-hop neighborhood of each agent would lead to biases on the order of $\gamma^{\kappa+1}$ in convergence.
	\end{remark}

	Let $\{M_i\}_{i=1}^N$ be solutions to (\ref{SDP grad}) for $i=1,...,N$. Also let $\hat{Q}_i=M_i\odot\tilde{Q}$ and  $\hat{R}_i=\sum_{j\in\mathcal{N}_L^i}(e_j^{\top}\otimes I_m)R_j(e_j\otimes I_m)$. By collecting the parts of the entire cost function involving agents in $\mathcal{N}_L^i$, we define the local cost $J_i$ as
		\begin{align}
			J_i(K)
			&=\mathbb{E}\left[\sum_{t=0}^\infty\gamma^t (x^{\top}\hat{Q}_ix+x^{\top}K^{\top}\hat{R}_iKx)\right]\notag\\
			&=\mathbb{E}\left[\sum_{t=0}^{\infty}\gamma^t(x_{\mathcal{N}_L^i}^{\top}\bar{Q}_ix_{\mathcal{N}_L^i}+ u_{\mathcal{N}_L^i}^{\top}\bar{R}_iu_{\mathcal{N}_L^i})
			\right],\label{local cost}
		\end{align}
	where  $\bar{Q}_i$ is the maximum nonzero principal submatrix of $M_i\odot\tilde{Q}$ and $\bar{R}_i=\diag\{R_j\}_{j\in\mathcal{N}_L^i}$. 
	
	\begin{remark}
		Each agent $i$ computes its local cost $J_i(x_{\mathcal{N}_L^i},u_{\mathcal{N}_L^i})$ based on $x_j$ and $u_j^{\top}R_ju_j$, $j\in\mathcal{N}_L^i$. In practical application, agent $i$ may compute a finite horizon cost value to approximate the infinite horizon cost. The state information $x_j$ is sensed from its neighbors in the learning graph, and $u_j^{\top}R_ju_j$ is obtained from its neighbors via communications. Note that $u_{\mathcal{N}_L^i}$ may involve state information of agents in $\mathcal{V}\setminus\mathcal{N}_L^i$. However, it is not necessary for agent $i$ to have such state information. Instead, it obtains $u_j^{\top}R_ju_j$ by communicating with agent $j\in\mathcal{N}_L^i\setminus\{i\}$.
	\end{remark}

	Next we verify the validity of Assumption \ref{as local cost}. Let $H_i(K,x(0))=\sum_{t=0}^\infty\gamma^t x^{\top}(t)(\hat{Q}_i+K^{\top}\hat{R}K)x(t)$. The next proposition shows that the local cost functions we constructed satisfy Assumption \ref{as local cost}.
	\begin{proposition}\label{pr LQR J_i}
		The local cost $J_i(K)$ constructed in (\ref{local cost}) has the following properties.
	
	(i). There exists a scalar $c_{lqr}>0$ such that $H_i(K,x(0))\leq c_{lqr}J_i(K)$ for any $K\in\mathbb{K}_s$ and $x(0)\sim\mathcal{D}$.
	
	(ii). $\nabla_{K_i}J_i(K)=\nabla_{K_i} J(K)$ for any $K\in\mathbb{K}_s$.
	\end{proposition}

	\vspace{-0.4cm}
	\subsection{Distributed RL Algorithms for Multi-Agent LQR}

	Due to Proposition \ref{pr LQR J_i}, we are able to apply Algorithms \ref{alg:as} to the distributed multi-agent LQR problem. Let $\mathbf{K}_i=\text{vec}(\tilde{K}_i)\in\mathbb{R}^{q_i}$, where $\tilde{K}_i$ is defined in (\ref{u_i}), $q_i=mn_i=mn|\mathcal{N}_S^i|$, $i=1,...,N$. From the degree sum formula, we know $$\sum_{i=1}^Nq_i=mn\sum_{i=1}^N|\mathcal{N}_S^i|=2mn|\mathcal{E}_S|\triangleq q.$$ 
	Define $\mathbf{K}=(\mathbf{K}_1^{\top},...,\mathbf{K}_N^{\top})^{\top}\in\mathbb{R}^{q}$. There must uniquely exist a control gain matrix $K\in\mathbb{R}^{mN\times nN}$ corresponding to $\mathbf{K}$. Let $\mathcal{M}_K(\cdot): \mathbb{R}^q\rightarrow\mathbb{K}_d$ be the mapping transforming $\mathbf{K}$ to a distributed stabilizing gain. 
	
	To apply Algorithm \ref{alg:as} to the multi-agent LQR problem, we need to implement Algorithm \ref{alg:clustering} based on the learning graph $\mathcal{G}_L$ to achieve a clustering $\{\mathcal{V}_j, j=1,...,s\}$. Then the asynchronous RL algorithm for problem (\ref{LQR no noise}) is given in Algorithm \ref{alg:lqr1}. Note that the algorithm requires a stabilizing distributed control gain as the initial policy. One way to achieve this is making each agent learn a policy stabilizing itself, which has been studied in \cite{Lamperski20}.
	
	To facilitate understanding of Algorithm \ref{alg:lqr1}, we present Table I to show the exact correspondence between the stochastic optimization (SO) problem in Section \ref{sec: as} and the LQR problem in this section. In Table I, $H(K,x(0))=\sum_{t=0}^\infty\gamma^tx^\top(t)(Q+K^\top RK)x(t)$. From Table I and the distributed learning diagram in Fig. \ref{fig diagram as}, we observe that the distributed learning nature of Algorithm \ref{alg:lqr1} is reflected by the learning graph $\mathcal{G}_L$. Specifically, each agent learns its policy based on its own estimated cost value and information from its neighbors in graph $\mathcal{G}_L$.

\begin{table}[H]
\caption{The correspondence between the SO problem and the LQR problem.}
\centering

\begin{tabular}{|c|c|c|c|c|c|c|}
	\hline Problem&\multicolumn{6}{|c|}{Variables}\\
	\hline SO&$x$&$\xi$&$h(x,\xi)$ & $f(x)$ & $T$ & $\mathcal{G}$\\
	\hline LQR&$\mathbf{K}$&$x(0)$&$H(K,x(0))$ & $J(K)$ & $T_K$ & $\mathcal{G}_L$\\
	\hline
\end{tabular}

\end{table}

	\begin{algorithm}[htbp]
		\small
		\caption{Asynchronous Distributed Learning for Multi-Agent LQR}\label{alg:lqr1}
		\textbf{Input}: Step-size $\eta$, smoothing radius $r_i$ and variable dimension $q_i$,  $i=1,...,N$, clusters $\mathcal{V}_j$, $j=1,...,s$, iteration numbers $T_K$ and $T_J$ for controller variable and the cost, respectively, update sequence $z_k$ and extrapolation weight $w_i^k$, $k=0,...,T_K-1$, $\mathbf{K}^0$ such that $\mathcal{M}_K(\mathbf{K^0})\in\mathbb{K}_s$.\\
		\textbf{Output}: $\mathbf{K}^*$.
		\begin{itemize}
			\item[1.] \textbf{for} $k=0,1,...,T_K-1$ \textbf{do}
			\item[2.] ~~~~~~~~Sample $x_0\sim\mathcal{D}$. Set $x(0)=x_0$.
			\item[3.]~~~~~~~~\textbf{for} $i\in\mathcal{V}$ \textbf{do} 
			\item[4.] ~~~~~~~~~~~~\textbf{if} $i\in\mathcal{V}_{z_k}$ \textbf{do} (Simultaneous Implementation)
			\item[5.] ~~~~~~~~~~~~~~~~Agent $i$ samples $\mathbf{D}_i(k)\in\mathbb{R}^{q_i}$ randomly from $\mathbb{S}_{q_i-1}$, and computes $\hat{\mathbf{K}}_i^k=\mathbf{K}_i^k+w_i^k(\mathbf{K}_i^k-\mathbf{K}_i^{k_{prev}})$.
			\item[6.] ~~~~~~~~~~~~~~~~Agent $i$ implements its controller $$u_i(t,k)=-\text{vec}^{-1}(\hat{\mathbf{K}}_i^k+r_i\mathbf{D}_i(k))x_{\mathcal{N}_S^i}(t),$$
			~~~~~~~~~~~~~~~~while each agent $j\in \mathcal{V}\setminus\mathcal{V}_{z_k}$ implements
			$u_{j}(t,k)=-\text{vec}^{-1}(\mathbf{K}_{j}^k)x_{\mathcal{N}_S^{j}}(t)$for $t=0,...,T_J-1$, and observes $H_{i,T_J}(K^{i,k},x(0))=$
			\begin{equation}
				\sum_{t=0}^{T_J-1}\gamma^t\left[x_{\mathcal{N}_L^i}^{\top}(t)\bar{Q}_ix_{\mathcal{N}_L^i}(t)+ u_{\mathcal{N}_L^i}^{\top}(t,k)\bar{R}_iu_{\mathcal{N}_L^i}(t,k)\right].\label{eq:finite_Cost}
			\end{equation}
			\item[7.] ~~~~~~~~~~~~~~~~Agent $i$ computes the estimated gradient: $$\mathbf{G}_i^{T_J}(k)=\frac{q_i}{r_i}H_{i,T_J}(K^{i,k},x(0))\mathbf{D}_i(k),$$
			~~~~~~~~~~~~~~~~then updates its policy:
			\begin{equation}
				\mathbf{K}_i^{k+1}=\hat{\mathbf{K}}_i^k-\eta \mathbf{G}_i^{T_J}(k).
			\end{equation}
		\item[8.]~~~~~~~~~~~~\textbf{else}
		\begin{equation}
			\mathbf{K}_i^{k+1}=\mathbf{K}_i^k.
		\end{equation}
		\item[9.] ~~~~~~~~~~~~\textbf{end}
		\item[10.]~~~~~~~~\textbf{end}
			\item[11.] \textbf{end}
			\item[12.] Set $\mathbf{K}^*=\mathbf{K}^{k+1}$.
		\end{itemize}
	\end{algorithm}	

 \vspace{-0.4cm}
	\subsection{Convergence Analysis}
	
	In this subsection, we show the convergence result of Algorithm \ref{alg:lqr1}. Throughout this subsection, we adopt $J_i(K)$ in (\ref{local cost}) as the local cost function for agent $i$. Let $Q_i^K=\hat{Q}_i+K^{\top}\hat{R}_iK$, $J_i^{T_J}(K)=\mathbb{E}[\sum_{t=0}^{T_J-1}x^{\top}(t)Q_i^Kx(t)]$, $\mathbf{G}=(\mathbf{G}_1^{\top},...,\mathbf{G}_N^{\top})^{\top}$, $\mathbf{G}_i(k)=\frac{q_i}{r_i}J_i(k)\mathbf{D}_i(k)$, $\mathbf{G}^{T_J}=((\mathbf{G}_1^{T_J})^{\top},...,(\mathbf{G}_N^{T_J})^{\top})^{\top}$, where $\mathbf{G}_i$ and $\mathbf{G}_i^{T_J}$ are the ideal and the actual estimates, respectively, of the gradient of agent $i$'s local cost function, and $\mathbf{D}_i(k)\in\mathbb{R}^{q_i}$ is randomly sampled from $\mathbb{S}_{q_i-1}$.

	The essential difference between Algorithm \ref{alg:as} and Algorithm \ref{alg:lqr1} is that the computation of the cost function in an LQR problem is inaccurate because~\eqref{eq:finite_Cost} only provides a finite horizon cost, whereas the cost in \eqref{local cost} without expectation is over an infinite horizon. This means that we need to take into account the estimation error of each local cost.

	Let $H_i(K,x(0))=\sum_{t=0}^{\infty}\gamma^t(x_{\mathcal{N}_L^i}^{\top}\bar{Q}_ix_{\mathcal{N}_L^i}+ u_{\mathcal{N}_L^i}^{\top}\bar{R}_iu_{\mathcal{N}_L^i})$, whose expectation is $J_i(K)$ in (\ref{local cost}). Then $H_i$ corresponds to $h_i$ in Algorithm \ref{alg:as}. Therefore, the ideal and actual gradient estimates for agent $i$ at step $k$ are  $\mathbf{G}_i(k)=\frac{q_i}{r_i}H_i(K^{i,k},x(0))\mathbf{D}_i(k)$, and $\mathbf{G}_i^{T_J}(k)=\frac{q_i}{r_i}H_{i,T_J}(K^{i,k},x(0))\mathbf{D}_i(k)$, respectively, where $K^{i,k}=\mathcal{M}_K(\mathbf{K}^{i,k})$, $\mathbf{K}^{i,k}=(\mathbf{K}_1^{k\top},...,(\hat{\mathbf{K}}_i^k+r_i\mathbf{D}_i(k))^{\top},...,\mathbf{K}_N^{k\top})^{\top}$. Note that for any $\mathbf{K}\in\mathbb{R}^q$, it holds that $\|\mathbf{K}\|=\|\mathcal{M}_K(\mathbf{K})\|_F$.
	
	We study the convergence by focusing on $K\in\mathbb{K}_\alpha$, where $\mathbb{K}_{\alpha}$ is defined as 
 	\begin{equation}\label{X}
		\mathbb{K}_\alpha=\{K\in\mathbb{K}_d: J(K)\leq\alpha J(K^0),J_i(K)\leq\alpha_iJ_i(K^0)\}\subseteq \mathbb{K}_s,
	\end{equation}
	where $\alpha,\alpha_i>1$, $K^0\in\mathbb{K}_d\cap\mathbb{K}_s$ is the given initial stabilizing control gain. Then $\mathbb{K}_\alpha$ is compact due to the coerciveness of $J(K)$ (as shown in Lemma \ref{lem:a2a3}).
	
	Due to the continuity of  $\phi_K$, $\lambda_K$, $\beta_K$ and $\zeta_K$, and the compactness of $\mathbb{K}_\alpha$, we define the following parameters for $K\in\mathbb{K}_\alpha$:
	\begin{equation}
		\begin{split}
\phi_0&=\sup_{K\in\mathbb{K}_\alpha}\phi_K,\lambda_0=\sup_{K\in\mathbb{K}_\alpha}\lambda_K, \rho_0=\inf_{K\in\mathbb{K}_\alpha}\{\beta_K,\zeta_K\},\\&~~\kappa_0=\sup_{K\in\mathbb{K}_\alpha}\frac{1+\rho(\sqrt{\gamma}(\mathcal{A}-\mathcal{B}K))}{2}.
	\end{split}
	\end{equation}
 
Motivated by \cite[Lemma 8]{Li19}, there exists a constant $C_K>0$ such that 
\begin{equation}\label{C_K}
    \|\left(\sqrt{\gamma}(\mathcal{A}-\mathcal{B}K)\right)^t\|\leq C_K\left(\frac{1+\rho(\sqrt{\gamma}(\mathcal{A}-\mathcal{B}K))}{2}\right)^t,
\end{equation}
where $C_K$ continuously depends on $\sqrt{\gamma}(\mathcal{A}-\mathcal{B}K)$. Reusing the compactness of $\mathbb{K}_\alpha$, we define $C_0=\sup_{K\in\mathbb{K}_\alpha}C_K$.

	The following lemma evaluates the gradient estimation in the LQR problem.
	
	\begin{lemma}\label{le LQR g as}
 Given $\epsilon'>0$ and $K^{k}\in\mathbb{K}_\alpha$,  if $r_i\leq \frac{\rho_0}{2}$, $w_i^k\leq\frac{\rho_0}{2\|K_i^k-K_i^{k_{prev}}\|_F}$ and
		\begin{equation}\label{T_J bound}
			T_J\geq\max_{i\in\mathcal{V}}\frac{1}{2(1-\kappa_0)}\log  \left(\frac{\alpha_iJ_i(K^0)C_0^2\|x(0)\|^2}{\lambda_{\min}(\Sigma_x)\epsilon'}\right),
		\end{equation}
		then
			\begin{equation}\label{J error}
			H_i(K^{i,k},x(0))-H_{i, T_J}(K^{i,k},x(0))\leq\epsilon', ~~~~i\in\mathcal{V},
		\end{equation}
		
		\begin{equation}\label{LQR g error}
			\|\mathbb{E}[\mathbf{G}_i^{T_J}(k)]-\nabla_{\mathbf{K}_i}J(K^{i,k})\|\leq \frac{q_i\epsilon'}{r_i}+\phi_0r_i,
		\end{equation}
	\begin{equation}\label{LQR bound g}
		\|G_i^{T_J}(k)\|\leq \frac{q_i}{r_i}\left[c_{lqr}(\alpha_i  J_i(K^0)+\lambda_0\rho_0)+\epsilon'\right],
	\end{equation}
		where $J_{i,T_J}$ is in (\ref{eq:finite_Cost}). 
 \end{lemma}

	Let $J_0(K^0)=\max_{i\in\mathcal{V}}\alpha_i J_i(K^0)$. Based on Lemma \ref{le LQR g as} and Theorem \ref{th as}, we have the following convergence result for Algorithm \ref{alg:lqr1}.
	\begin{theorem}\label{th as LQR}
		Under Assumption \ref{as solvable}, given positive scalars $\epsilon$, $\nu$, $\gamma$, and $\alpha\geq 2+\gamma+\frac{1}{\nu}+\nu\gamma$, $K^0\in\mathbb{K}_s\cap\mathbb{K}_d$. Let $\{\mathbf{K}^k\}_{k=0}^{T_K-1}$ be the sequence of states obtained by implementing Algorithm \ref{alg:lqr1} for $k=0,...,T_K-1$. Suppose that $T_J$ satisfies (\ref{T_J bound}) for each $k$ ($T_J$ can be different for different steps), and
		\begin{equation}\label{lqrconditions}
		\begin{split}
		    &T_K=\lceil \frac{2\alpha \nu J(\mathbf{K}^0)}{\eta\epsilon}\rceil,~~~ r_i\leq \min\{\frac{\rho_0}{2},\frac{1}{4\phi_0}\sqrt{\frac{\gamma\epsilon}{\alpha N_0}}\},\\ &\eta\leq\min\{\frac{\rho_0}{2\delta\sqrt{N_0}},\frac{2\alpha J(\mathbf{K}^0)}{\gamma\epsilon},\frac{\gamma\epsilon}{2\alpha N_0(\phi_0\delta_2^2+4\phi_0^2+\phi_0+4)}\},\\ &\epsilon'\leq\frac{r_-}{4q_+}\sqrt{\frac{\gamma\epsilon }{\alpha N_0}}, 
		    w_i^k\leq\frac{1}{\|\mathbf{K}_i^k-\mathbf{K}_i^{k_{prev}}\|}\min\{\eta^{3/2},\frac{\rho_0}{2\sqrt{N_i}}\},
		    \end{split}
		\end{equation}
	for $i\in\mathcal{V}$,	where $\delta=\frac{q_+}{r_-}\left[c_{lqr}(\alpha_i J_0(K^0)+\lambda_0\rho_0)+\epsilon'\right]$.
		
	(i). The following holds with a probability at least $1-\frac{1}{\alpha}(2+\gamma+\frac{1}{\nu}+\nu\gamma)$:
		\begin{equation}
			\frac{1}{T_K}\sum_{k=0}^{T_K-1}\sum_{i\in\mathcal{V}_{z_k}}\|\nabla_{\mathbf{K}_i}J(\mathbf{K}^k)\|^2< \epsilon.
		\end{equation}		
	(ii). Under Assumption \ref{as cyclic}, suppose that there exists some $\bar{\epsilon}\in(0,\rho_0]$ such that
	\begin{equation}
	w_i^k\leq \frac{\bar{\epsilon}}{2(T_0-1)N_i\|\mathbf{K}_i^k-\mathbf{K}_i^{k_{prev}}\|},~~ \eta\leq\frac{\bar{\epsilon}}{2\delta (T_0-1)N_0},
	\end{equation}
	for any step $k\in\{0,..., T_K-1\}$, and conditions in (\ref{lqrconditions}) are satisfied,  then the following holds with a probability at least $1-\frac{1}{\alpha}(2+\gamma+\frac{1}{\nu}+\nu\gamma)$:
	\begin{equation}
	    \frac{1}{T_K}\sum_{k=0}^{T_K-1}\|\nabla_\mathbf{K}J(\mathbf{K}^k)\|^2<\hat{\epsilon} .
	\end{equation}
	where $\hat{\epsilon}=2T_0(\epsilon+\phi_0^2\bar{\epsilon}^2)$.
	\end{theorem}
	
	From Theorem \ref{th as LQR} (ii), the sample complexity for convergence of Algorithm \ref{alg:lqr1} is $T_KT_J=\mathcal{O}(q_+^2N_0^2\log(q_+/\hat{\epsilon})/\hat{\epsilon}^3)$, which is higher than that of Algorithm \ref{alg:as} because of the error on the local cost function evaluation in the LQR problem. The sample complexity here has a lower order on convergence accuracy than that in \cite{Li19}. Moreover, our algorithm has another two advantages: (i) the sample complexity in \cite{Li19} is affected by the convergence rate of the consensus algorithm, the number of agents, and the dimension of the entire state variable, while the sample complexity of our algorithm depends on the local optimization problem for each agent (the number of agents in one cluster and the dimension of the variable for one agent), rendering our algorithm high scalability to large-scale networks; (ii) the algorithm in \cite{Li19} requires each agent to estimate the global cost during each iteration, while our algorithm is based on local cost evaluation, which benefits for variance reduction and privacy preservation.

	\section{Simulation Experiments}\label{sec: sim}	
	
	\subsection{Optimal Tracking of Multi-Robot Formation}
\begin{figure}
		\centering
		\includegraphics[width=9cm]{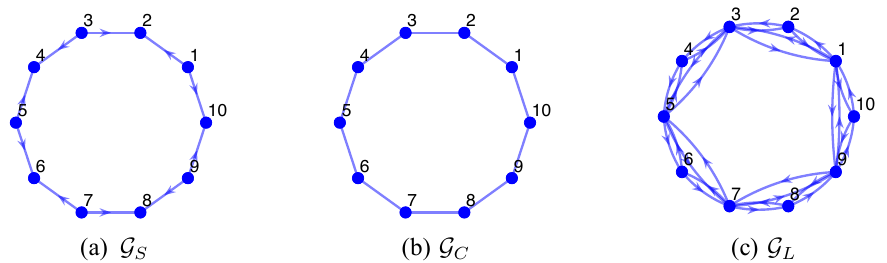}
		\caption{The sensing graph $\mathcal{G}_S$, cost graph $\mathcal{G}_C$ and the resulting learning graph $\mathcal{G}_L$ for the formation control problem.} \label{fig formation graphs}
	\end{figure}

	In this section, we apply Algorithm \ref{alg:lqr1} to a multi-agent formation control problem. Consider a group of  $N=10$  robots modeled by the following double integrator dynamics:
	\begin{equation}\label{agent dynamics}
		\begin{split}
			r_i(t+1)&=r_i(t)+v_i(t),\\
			v_i(t+1)&=v_i(t)+C_iu_i(t), ~~i=1,...,10,
		\end{split}
	\end{equation}
	where $r_i, v_i, u_i\in\mathbb{R}^2$ are position, velocity, and control input of agent $i$, respectively, $C_i\in\mathbb{R}^{2\times2}$ is a coupling matrix in the dynamics of agent $i$. Let $x_i=(r_i^{\top},v_i^{\top})^{\top}$, $A_i=(\begin{smallmatrix}
		I_2&I_2\\
		\mathbf{0}_{2\times2}&I_2
	\end{smallmatrix})$, $B_i=(\mathbf{0}_{2\times2},C_i^{\top})^{\top}$, the dynamics (\ref{agent dynamics}) can be rewritten as
	\begin{equation}
		x_i(t+1)=A_ix_i(t)+B_iu_i(t).
	\end{equation}
	The control objective is to make the robots learn their own optimal controllers for the whole group to form a circular formation, track a moving target, maintain the formation as close as possible to the circular formation during the tracking process, and cost the minimum control energy. The target has the following dynamics:
	\begin{equation}
		r_0(t+1)=r_0(t)+v_0,
	\end{equation}
	where the velocity $v_0\in\mathbb{R}^2$ is fixed. Let $x_0=(r_0^{\top},v_0^{\top})^{\top}$ be the state vector of the target, and $d_i(t)=x_0(t)+(\cos\theta_i,\sin\theta_i,0,0)^{\top}$ with $\theta_i=\frac{2\pi i}{N}$ be the desired time-varying state of robot $i$. Suppose that the initial state $x_i$ of each robot $i$ is a random variable with mean $d_i$, which implies that each agent is randomly perturbed from its desired state. Then the objective to be minimized can be written as 	
 \begin{equation*}
 \begin{split}
	J=&\mathbb{E}_{(x(0)-d)\sim\mathcal{D}}\Big[\sum_{t=0}^\infty \Big(\sum_{(i,j)\in\mathcal{E}_C}\|x_i(t)-x_j(t)-(d_i-d_j)\|^2\\		
&+\sum_{i\in\mathcal{V}_r}\|x_i-d_i\|^2
		+\sum_{i=1}^N\|u_i(t)\|^2\Big)\Big],
  \end{split}
  \end{equation*}
where $\mathcal{V}_r$ is the leader set consisting of robots that do not sense information from others but only chase for its target trajectory. In the literature of formation control, usually the sensing graph via which each robot senses its neighbors' state information is required to have a spanning tree. Here we assume that the sensing graph has a spanning forest from the leader set $\mathcal{V}_L$. Different from a pure formation control problem, the main goal of the LQR formulation is to optimize the transient relative positions between different robots and minimize the required control effort of each agent. Fig. \ref{fig formation graphs} shows the sensing graph, the cost graph and the resulting learning graph for the formation control problem. The leader set is $\{1,3,5,7,9\}$.

	Define the new state variable $y_i=x_i-d_i$ for each robot $i$. Using the property $A_id_i(t)=d_i(t+1)$, we obtain dynamics of $y_i$ as $y_i(t+1)=A_iy_i(t)+B_iu_i(t)$. The objective becomes
	\begin{equation*}
		\min_{u} J=\mathbb{E}_{y(0)\sim\mathcal{D}}\left[\sum_{t=0}^\infty \gamma^t(y^{\top}(t)(L+\Lambda)y(t)+u^{\top}(t)u(t))\right],
	\end{equation*}
	where $y=(y_1^{\top},...,y_N^{\top})^{\top}$, $L$ is the Laplacian matrix corresponding to the cost graph, $\Lambda$ is a diagonal matrix with $\Lambda_{ii}=1$ if $i\in\mathcal{V}_r$ and $\Lambda_{ii}=0$ otherwise.

	Let $C_i=\frac{i}{i+1}I_2$, the initial stabilizing gain is given by $\mathbf{K}(0)=\mathcal{M}_K^{-1}(K_0)$, where $K_0=I_N\otimes \tilde{K}$ and $\tilde{K}=(I_2,1.5I_2)$, which is stabilizing for each robot. By implementing Algorithm \ref{alg:clustering} based on the learning graph in Fig. \ref{fig formation graphs} (c),  the following clustering is obtained :
	\begin{equation}\label{formation cluster}
		\mathcal{V}_1=\{1,4,6,8\}, \mathcal{V}_2=\{2,5,9\}, \mathcal{V}_3=\{3,7,10\},
	\end{equation}
	which means that there are three clusters update asynchronously, while agents in each cluster update their variable states via independent local cost evaluations at one step. It will be shown that compared with the  traditional BCD algorithm where only one block coordinate is updated at one step, our algorithm is more efficient.
	
	\subsection{Simulation Results}
	\begin{figure*}
		\centering
		\includegraphics[width=18cm]{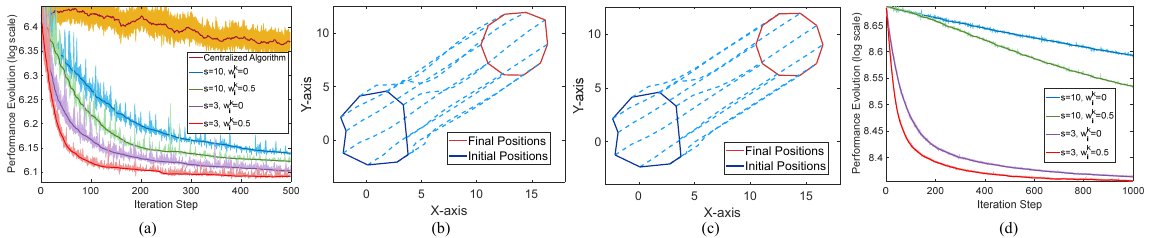}
		\caption{(a) The group performance evolution of a 10-agent formation under the centralized ZOO algorithm, Algorithm \ref{alg:lqr1} without clustering and acceleration, without clustering but with acceleration, with clustering but without acceleration, and with both clustering and acceleration. The shaded areas denote the performance corresponding to the controllers obtained by perturbing the current control gain with 50 random samplings. (b) The trajectories of robots under the initial stabilizing controller. (c) The trajectories of robots under the controller learned by Algorithm \ref{alg:lqr1} with $s=3$, $w_i^k=0.5$. (d) The performance evolution of a 100-agent formation under Algorithm \ref{alg:lqr1}.} \label{fig simulations}
	\end{figure*}
	 We compare our algorithm with the centralized one-point ZOO algorithm\footnote{There have been many distributed RL algorithms under different settings in recent years. Distributed algorithms with consensus-based global information estimation (e.g., \cite{Li19,Zhang2018fully}) are  applicable to the problem solved in this paper. However, they usually converge slower than their centralized counterparts due to the consensus-based estimation/interaction. Therefore, we only compare with the centralized algorithm to show our improved convergence rate.} in \cite{Fazel18}, where $\mathbf{K}$ is updated by taking an average of multiple repeated global cost evaluations. Fig. \ref{fig simulations} shows the simulation results, where the performance is evaluated with given initial states of all the robots.  When implementing the algorithms, each component of the initial states are randomly generated from a truncated normal distribution with mean 0 and variance 1  . Moreover, we set $\eta=10^{-6}$, $r=1$, $T_K=1000$, $T_J=50$ for both Algorithm \ref{alg:lqr1} and the centralized algorithm. In Fig. \ref{fig simulations}, we show the performance trajectory of the controller generated by different algorithms. When $s=10$, Algorithm \ref{alg:lqr1} is a BCD algorithm without clustering, i.e., each cluster only contains one robot, while $s=3$ corresponds to the clustering strategy in (\ref{formation cluster}). The extrapolation weight $w_i^k$ for each agent $i$ at each step $k$ is uniformly set. So $w_i^k=0$ implies that Algorithm \ref{alg:lqr1} is not accelerated. To compare the variance of the controller performance for different algorithms, we conduct gradient estimation for 50 times in each iteration, and plot the range of performance induced by all the estimated gradients, as shown in the shaded areas of Fig. \ref{fig simulations} (a). Each solid trajectory in Fig. \ref{fig simulations} corresponds to a controller updated by using an average of the estimated gradients for the algorithm. Fig. \ref{fig simulations} (b) and (c) show the trajectories of agents by implementing the initial stabilizing controller and the convergent controller learned by Algorithm \ref{alg:lqr1} with $s=3$, and $w_i^k=0.5$, $i\in\mathcal{V}$, $k=0,...,T_K-1$, respectively, for $t\in[0,15]$. From the simulation results, we have the following observations:
	 
	 \begin{itemize}
	     \item Algorithm \ref{alg:lqr1} always converges faster and has a lower performance variance than the centralized zeroth-order algorithm.
      \item The learned controller achieves the formation tracking faster than the initial controller, therefore has a better performance.
	     \item Algorithm \ref{alg:lqr1} with the cluster-wise update scheme converges faster than it with the agent-wise update scheme. This is because the number of clusters dominates the sample complexity, as we analyzed below Theorem \ref{th as LQR}.
	     \item Appropriate extrapolation weights not only accelerate Algorithm \ref{alg:lqr1}, but also decrease the performance variance.
	 \end{itemize}
	
	Note that the performance of the optimal centralized controller is 6.03 (in log scale), which is close to the performance of the learned control gain\footnote{The reason why Algorithm \ref{alg:lqr1} converges to a controller with a cost value larger than the optimal one in Fig. \ref{fig simulations} (a) is that the distributed LQR problem has a structure constraint for the control gain, which results in a feasible set $\mathbb{K}_s\cap\mathbb{K}_d$, whereas the feasible set of the centralized LQR problem is $\mathbb{K}_s$.}. By conducting ten more experiments with different initial stabilizing gains, we find that the performance of the learned control gain does not vary much. The accurate sub-optimality cannot be estimated since it is difficult to obtain the globally optimal structured control gain.
	
	\textbf{Scalability to Large-Scale Networks.} Even if we further increase the number of  robots, the local cost of each robot always involves only 5 robots as long as the structures of the sensing and cost graphs are maintained. That is, the magnitude of each local cost does not change as the network scale grows. On the contrary, for global cost evaluation, the problem size severely influences the estimation variance. The difference of variances between these two methods has been analyzed in Subsection \ref{subsec: as variance}. To show the advantage of our algorithm, we further simulate a case with 100 robots, where the cost, sensing and learning graphs have the same structure as the 10-robot case, and implementing Algorithm \ref{alg:clustering} still results in 3 clusters. Simulation results show that the centralized algorithm failed to solve the problem. The performance trajectories of Algorithm \ref{alg:lqr1} with different settings are shown in Fig. \ref{fig simulations} (c), from which we observe that Algorithm \ref{alg:lqr1} has an excellent performance, and both clustering and extrapolation are efficient in improving the convergence rate. Note that the performance of the centralized optimal control gain for the 100-agent case is $\log(4119.87)=8.3236$, which is very close to the performance of the learned structured controller.

	\section{Conclusion}\label{sec: con}
	We have proposed a novel distributed RL (zeroth-order accelerated BCD) algorithm with asynchronous sample and update schemes. The algorithm was applied to distributed learning of the model-free distributed LQR by designing the specific local cost functions and the interaction graph required by learning. A simulation example of formation control has shown that our algorithm has significantly lower variance and faster convergence rate compared with policy gradient algorithms with global value function evaluation. In the future, we will look into how to reduce the sample complexity of the proposed distributed ZOO algorithm.

	\section{Appendix}\label{sec: app}

\subsection{An Algorithm for Non-Adjacent Agents Clustering}\label{app alg} 

Algorithm \ref{alg:clustering} provides an approach\footnote{Two notes regarding Algorithm \ref{alg:clustering} are given here. Note 1: the number of clusters obtained by implementing Algorithm \ref{alg:clustering} may be different each time. In practical applications, Algorithm \ref{alg:clustering} can be modified to solve for a clustering with maximum or minimum number of clusters. Note 2: Algorithm \ref{alg:clustering} requires the global graph information as an input, and therefore, is a centralized algorithm. However, Algorithm \ref{alg:clustering} is only implemented in one shot, and there is no real-time global information required during its implementation. Therefore, Algorithm \ref{alg:clustering} can be viewed as an off-line centralized deployment before implementing the distributed policy seeking algorithm.} for decomposing the agents into $s$ independent clusters $\{\mathcal{V}_j\}_{j=1}^s$ such that $(i_1,i_2)\notin\mathcal{E}$ for any $i_1,i_2\in\mathcal{V}_{j}$, $j\in\{1,...,s\}$.		
	\begin{algorithm}[htbp!]
		\small
		\caption{Non-Adjacent Agents Clustering}\label{alg:clustering}
		\textbf{Input}: $\mathcal{V}$, $\mathcal{N}_i$ for $i=1,...,N$.\\
		\textbf{Output}: $s$, $\mathcal{V}_j$, $j=1,...,s$.
		\begin{itemize}
			\item[1.] Set $s=0$, $\mathcal{C}=\varnothing$.
			\item[2.] \textbf{while} $\mathcal{C}\neq\mathcal{V}$
			\item[3.]  Set $s\leftarrow s+1$, $\mathcal{D}=\mathcal{C}$, \textbf{while} $\mathcal{D}\neq\mathcal{V}$
			\item[4.] Randomly select $i$ from $\mathcal{V}\setminus\mathcal{D}$, set $\mathcal{V}_s\leftarrow\mathcal{V}_s\cup\{i\}$, $\mathcal{D}\leftarrow\mathcal{D}\cup\mathcal{N}_i$.
			\item[5.] \textbf{end} 
			\item[6.] $\mathcal{C}\leftarrow\mathcal{C}\cup\mathcal{V}_s$.
			\item[7.] \textbf{end}
	\end{itemize}
	\end{algorithm}

	\subsection{Analysis for the Asynchronous RL Algorithm}

	{\it Proof of Lemma \ref{le fhat=Eg}: } It suffices to show that	
	\begin{equation}\label{estimate for gi}
		\mathbb{E}_{u_i^l\in\mathbb{S}_{n-1}}[f_i(x_i+r_i u_i^l,x_{-i})u_i^l]=\frac{r_i}{q_i}\nabla_{x_i}\hat{f}_i(x),
	\end{equation}
	for $i=1,...,N$.
	This has been proved in \cite[Lemma 1]{Flaxman04}.
	\QEDA

	{\it Proof of Lemma \ref{le fhat error}:} Denote $v^i=(0,...,0,v_i^{\top},0,...,0)^{\top}\in\mathbb{R}^q$ with $v_i\in\mathbb{R}^{q_i}$. The following holds: 
	\begin{equation}
		\begin{split}
			&\|\nabla_{x_i}\hat{f}_i(x)-\nabla_{x_i}f(x)\|\\
			&=\|\nabla_{x_i}\mathbb{E}_{v_i\in\mathbb{B}_{q_i}}[f_i(x_i+r_i v_i,x_{-i})]-\nabla_{x_i}f(x)\|\\
			&=\|\mathbb{E}_{v_i\in\mathbb{B}_{q_i}}\left[\nabla_{x_i}f_i(x_i+r_i v_i,x_{-i})-\nabla_{x_i}f(x)\right]\|\\
			&\leq \mathbb{E}_{v_i\in\mathbb{B}_{q_i}}\|\nabla_{x_i}f_i(x_i+r_i v_i,x_{-i})-\nabla_{x_i}f(x)\|\\
			&=\mathbb{E}_{v^i\in\mathbb{B}_{q}}\|\nabla_{x_i}f(x+r_i v^i)-\nabla_{x_i}f(x)\|\\
			&\leq \mathbb{E}_{v^i\in\mathbb{B}_{q}}\|\nabla_{x}f(x+r_i v^i)-\nabla_{x}f(x)\|\\
			&\leq \phi_xr_i,
		\end{split}
	\end{equation}
	where the second equality follows from the smoothness of $f_i(x)$, and the first inequality used Jensen's inequality.
	\QEDA

Before proving Theorem \ref{th as}, we show that once $x^k$ is restricted in $\mathbb{X}$, we are able to give a uniform bound on the estimated gradient $g_i(\hat{x}_i,x_{-i}^k, u,\xi)$ at each step, see the following lemma.
	
	\begin{lemma}\label{le E[g]}
		If  $r_i\leq \rho_0/2$ and $w_i^k\leq \frac{\rho_0}{2\|x_i^k-x_i^{prev}\|}$, then for any $x^k\in\mathbb{X}$, the estimated gradient satisfies
		\begin{equation}
			\|g_i(\hat{x}_i^k,x_{-i}^k,u_i,\xi)\|\leq\frac{q_i}{r_i}c\left[\alpha_i f_i(x^0)+\lambda_0\rho_0\right].
		\end{equation}
	\end{lemma}
	
	{\it Proof:} Using the definition of $g_i(x,u_i,\xi)$ in (\ref{gixuxi}), we have
	\begin{align}
		&\|g_i(\hat{x}_i^k,x_{-i}^k,u_i,\xi)\|
		=\frac{q_i}{r_i}h_i(\hat{x}_i^k+r_iu_i,x_{-i}^k,\xi) \notag\\
		&\leq\frac{q_i}{r_i}cf_i(\hat{x}_i^k+r_iu_i,x_{-i}^k) \leq\frac{q_i}{r_i}c\left[f_i(x^k)+\lambda_0\rho_0\right] \notag\\
		&\leq\frac{q_i}{r_i}c\left[\alpha_i f_i(x^0)+\lambda_0\rho_0\right],\notag
	\end{align}
where the first inequality used Assumption \ref{as local cost}, the second inequality used the local Lipschitz continuity of $f(x)$ over $\mathbb{X}$ because
\begin{equation*}
    \|\hat{x}_i^k+r_iu_i-x_i^k\|\leq\|\hat{x}_i^k-x_i^k\|+r_i=w_i^k\|x_i^k-x_i^{k_{prev}}\|+r_i\leq\rho_0,
\end{equation*}
and the last inequality holds because $x^k\in\mathbb{X}$.
\QEDA	
	
	{\it Proof of Theorem \ref{th as}:} To facilitate the proof, we give some notations first. Let $\mathcal{K}_i^k=\{j: i\in\mathcal{V}_{z_j}, j\leq k\}$ be the set of iterations that $x_i$ is updated before step $k$,  $d_i^k=|\mathcal{K}_i^k|$ be the number of times that $x_i$ has been updated until step $k$. Let $\tilde{x}_i^j$ be the value of $x_i$ after it is updated $j$ times, then $x_i^k=\tilde{x}^{d_i^k}_i$. We use $g_i(\hat{x}_i^k,x_{-i}^k)$ as the shorthand of $g_i(\hat{x}_i^k,x_{-i}^k,u,\xi)$. Let $\mathcal{F}_k$ denote the $\sigma$-field containing all the randomness in the first $k$ iterations. Moreover, we use the shorthand $\mathbb{E}^k[\cdot]=\mathbb{E}[\cdot|\mathcal{F}_k]$.

(i). Suppose $x_k\in\mathbb{X}$. Since $f(x)$ has a $(\phi_0,\rho_0)$ Lipschitz continuous gradient at $x_k$,
and 
\begin{align}
    \|x^{k+1}-x^k\|^2&=\sum_{i\in\mathcal{V}_{z_k}}\|x_i^{k+1}-x_i^k\|^2\nonumber\\
    &\leq \sum_{i\in\mathcal{V}_{z_k}}\left[w_i^k(x_i^k-x_i^{k_{prev}})+\eta\|g_i(\hat{x}_i^k,x_{-i}^k)\|\right]^2\nonumber\\
    &\leq N_i(\frac{\rho_0}{2\sqrt{N_i}}+\frac{\rho_0}{2\sqrt{N_0}})^2 \leq\rho_0^2,
\end{align}
we have
	\begin{align}
			&f(x^{k+1})-f(x^k)\nonumber\\
			&\leq \sum_{i\in\mathcal{V}_{z_k}}\left[\langle\nabla_{x_i}f(x^k),x^{k+1}_i-x^k_i\rangle +\frac{\phi_0}{2}\|x^{k+1}_i-x^k_i\|^2\right]\nonumber\\
			&\leq\sum_{i\in\mathcal{V}_{z_k}}\Big[\langle\nabla_{x_i} f_i(x^k),w_i^k(x_i^k-x_i^{k_{prev}})-\eta g_i(\hat{x}_i^k, x_{-i}^k)\rangle\nonumber\\
			&+\frac{\phi_0}{2}\eta^2\|g_i(\hat{x}_i^k,x_{-i}^k)\|^2
			+\frac{\phi_0}{2}(w_i^k)^2\|x_i^k-x_i^{k_{prev}}\|^2\Big]\nonumber\\
			&=\sum_{i\in\mathcal{V}_{z_k}}\Big[-\eta\|\nabla_{x_i} f_i(x^k)\|^2+\frac{\phi_0}{2}\eta^2\|g_i(\hat{x}_i^k,x_{-i}^k)\|^2\nonumber\\
			&~~~~+\eta\Delta_i^k+\Theta_i^k\Big],\label{fxk+1-fxk}
	\end{align}
	where $\Delta_i^k=\|\nabla_{x_i}f_i(x^k)\|^2-\langle\nabla_{x_i} f_i(x^k),g_i(\hat{x}_i^k,x_{-i}^k)\rangle$, $\Theta_i^k=\langle\nabla_{x_i}f(x^k), w_i^k(x_i^k-x_i^{k_{prev}})\rangle+\frac{\phi_0}{2}(w_i^k)^2\|x_i^k-x_i^{k_{prev}}\|^2$.
	
	Define the first iteration step at which $x$ escapes from $\mathbb{X}$ as
	\begin{equation}
		\tau=\min\{k: x(k)\notin\mathbb{X}\}.
	\end{equation}
	Next we analyze $\mathbb{E}^k\left[(f(x^k)-f(x^{k+1}))1_{\tau>k}\right]$. Under the condition $\tau>k$, both $\|\nabla_{x_i} f_i(\hat{x}_i^k,x_{-i}^k)-\mathbb{E}^k[g_i(\hat{x}_i^k,x_{-i}^k)]\|$ and $\|g_i(\hat{x}_i^k,x_{-i}^k)\|$ are uniformly bounded. For notation simplicity in the rest of the proof, according to Lemma \ref{le fhat error} and Lemma \ref{le E[g]}, we adopt $\theta=\phi_0\max_{i\in\mathcal{V}}r_i$ as the uniform upper bound of 
	$\|\nabla_{x_i} f_i(\hat{x}_i^k,x_{-i}^k)-\mathbb{E}^k[g_i(\hat{x}_i^k,x_{-i}^k)]\|$, and $\delta=\frac{q_+}{r_-}c\left[\alpha f_0(x^0)+\lambda_0\rho_0\right]$ as the uniform upper bound of $\|g_i(\hat{x}_i^k,x_{-i}^k)\|$.
	
	According to Assumption \ref{as local cost}, $\nabla_{x_i}f_i(x)$ is $(\phi_0,\rho_0)$ locally Lipschitz continuous. Then we can bound $\mathbb{E}^k[\Delta_i^k]$ if $\tau>k$:
	\begin{align}
			&\mathbb{E}^k[\Delta_i^k]=\langle\nabla_{x_i} f_i(x^k),\nabla_{x_i} f_i(x^k)-\nabla_{x_i} f_i(\hat{x}_i^k,x_{-i}^k)\rangle\nonumber\\
			&+\langle\nabla_{x_i} f_i(x^k),\nabla_{x_i} f_i(\hat{x}_i^k,x_{-i}^k)-\mathbb{E}^k[g_i(\hat{x}_i^k,x_{-i}^k)]\rangle\nonumber\\
			&\leq \|\nabla_{x_i} f_i(x^k)\|\phi_0\|x_i^k-\hat{x}_i^k\|+\theta \|\nabla_{x_i} f_i(x^k)\|\nonumber\\
			&=w_i^k\phi_0\|\nabla_{x_i} f_i(x^k)\|\|\tilde{x}_i^{d_i^k}-\tilde{x}_i^{d_i^k-1}\|+\theta \|\nabla_{x_i} f_i(x^k)\|\nonumber\\
			&\leq \frac{1}{8}\|\nabla_{x_i}f_i(x^k)\|^2+2(w_i^k)^2\phi_0^2\|\tilde{x}_i^{d_i^k}-\tilde{x}_i^{d_i^k-1}\|^2+\theta \|\nabla_{x_i} f_i(x^k)\|\nonumber\\
			&\leq \frac{3}{8}\|\nabla_{x_i}f_i(x^k)\|^2+2\phi_0^2\eta^3+\theta^2,\label{BoundDelta}
 \end{align}
	where the first inequality used $\|x_i^k-\hat{x}_i^k\|=\|w_i^k(x_i^k-x_i^{k_{prev}})\|\leq\rho_0$, and the Lipschitz continuity of $\nabla_{x_i}f(x_i^k,x_{-i}^k)$, the second equality used (\ref{xhati}), the second and the last inequalities used Young's inequality, and $w_i^k\leq\eta^{3/2}/\|x_i^k-x_i^{k_{prev}}\|$.

    Similarly, we bound $\mathbb{E}^k[\Theta_i^k]$ for $\tau>k$ as follows.
    \begin{equation}\label{BoundTheta}
        \mathbb{E}^k[\Theta_i^k]\leq \frac18\eta\|\nabla_{x_i}f_i(x^k)\|^2+2\eta^2+\frac{\phi_0}{2}\eta^3,
    \end{equation}
	where we used $w_i^k\leq\frac{\eta^{3/2}}{\|x_i^k-x_i^{k_{prev}}\|}$.


	Combining the inequalities (\ref{fxk+1-fxk}), (\ref{BoundDelta}) and (\ref{BoundTheta}), we have
	\begin{equation}\label{fk+1-fk}
		\begin{split}
			&\mathbb{E}^k[(f(x^{k+1})-f(x^k))1_{\tau>k}]\\
			&\leq-\frac12\sum_{i\in\mathcal{V}_{z_k}}\eta\mathbb{E}^k\left[\|\nabla_{x_i}f(x^k)\|^21_{\tau>k}\right]+\frac{\eta Z}{2},
		\end{split}
	\end{equation}	
	where $Z=N_0(\phi_0\delta^2\eta+4\phi_0^2\eta^3+2\theta^2+4\eta+\phi_0\eta^2)$.	
	
	Note that
	\begin{equation*}
		\begin{split}
\mathbb{E}^k\left[f(x^{k+1})1_{\tau>k+1}\right]&\leq\mathbb{E}^k \left[f(x^{k+1})1_{\tau>k}\right]= \mathbb{E}^k\left[f(x^{k+1})\right] 1_{\tau>k},
		\end{split}
	\end{equation*}
where the equality holds because all the randomness before the $(k+1)$th iteration has been considered in $\mathbb{E}^k[\cdot]$. Then $1_{\tau>k}$ is determined.
	It follows that
	\begin{multline}\label{succ}
		\mathbb{E}^k\left[(f(x^k)-f(x^{k+1}))1_{\tau>k}\right]
		\\ \leq\mathbb{E}^k\left[f(x^k)1_{\tau>k}-f(x^{k+1})1_{\tau>k+1}\right].
	\end{multline}	
	
	Summing (\ref{fk+1-fk}) over $k$ from $0$ to $T-1$ and utilizing (\ref{succ}) yields
	\begin{equation}
		\begin{split}
			&\mathbb{E}^k[f(x^{\top})1_{\tau>T}]-f(x^0)\\
			&\leq-\frac12\sum_{k=0}^{T-1}\sum_{i\in\mathcal{V}_{z_k}}\eta\mathbb{E}^k\left[\|\nabla_{x_i}f(x^k)\|^2\right]+T\eta Z/2.
		\end{split}
	\end{equation}
	It follows that
	\begin{equation}\label{gradnorm}
		\frac1T\sum_{k=0}^{T-1}\sum_{i\in\mathcal{V}_{z_k}}\mathbb{E}^k[\|\nabla_{x_i}f(x^k)\|^2]\leq \frac2{\eta T}f(x^0)+ Z.
	\end{equation}

	Now we analyze the probability $\mathbf{P}(\tau< T)$. Define the process
	\begin{equation}
		Y(k)=f(x^{\min\{k,\tau\}})+\frac{\eta}{2}(T-k)Z, ~~k=0,...,T-1,
	\end{equation}
	which is non-negative and almost surely bounded under the given conditions. Next we show $Y(k)$ is a supermartingale by considering the following two cases.
	\textit{Case 1: $\tau>k$.} 	
 \begin{equation}
		\begin{split}	\mathbb{E}^k[Y(k+1)]&=\mathbb{E}^k[f(x^{k+1})]+\frac{\eta}{2}(T-k-1)Z\\
			= f(x^k)+&\mathbb{E}^k[f(x^{k+1})-f(x^k)]+\frac{\eta}{2}(T-k-1)Z\\
			\leq f(x^k)+&\frac{\eta}{2}(T-k)Z=Y(k),
		\end{split}
	\end{equation}
	where the inequality used (\ref{fk+1-fk}).
	\textit{Case 2: $\tau\leq k$.}	
 \begin{equation}
 \begin{split}\mathbb{E}^k[Y(k+1)]&=f(x^\tau)+\frac{\eta}{2}(T-k-1)Z\\
 &\leq f(x^\tau)+\frac{\eta}{2}(T-k)Z= Y(k). 
 \end{split}
	\end{equation}
Therefore, $Y(k)$ is a super-martingale. Invoking Doob's maximal inequality for  super-martingales yields
	\begin{equation}
		\begin{split}
			\mathbf{P}(\tau< T) &\leq\mathbf{P}(\max_{k=0,...,T-1}f(x^k)>\alpha f(x^0))\\&\leq\mathbf{P}(\max_{k=0,...,T-1}Y(k)>\alpha f(x^0))\\
			&\leq\frac{\mathbb{E}^k[Y(0)]}{\alpha f(x^0)}=\frac{f(x^0)+\eta TZ/2}{\alpha  f(x^0)}=\frac{2+\nu\gamma}{\alpha}.
		\end{split}
	\end{equation}
	where the last equality used $\eta\leq\frac{2\alpha f(x^0)}{\gamma\epsilon}$, $T\leq\frac{2\alpha f(x^0)}{\epsilon\eta}\nu+1$, and $Z\leq \gamma\epsilon/\alpha$. The upper bound for $Z$ is obtained by noting that our conditions on $r_i$ and $\eta$ cause that $2\theta^2\leq\frac{\gamma\epsilon}{2\alpha N_0}$ and $\phi_0\delta^2\eta+4\phi_0^2\eta^3+4\eta+\phi_0\eta^2\leq(\phi_0\delta_2^2+4\phi_0^2+4+\phi_0)\eta\leq \frac{\gamma\epsilon}{2\alpha N_0}$ (we consider $\eta\leq1$ by default).
	
	It follows that 
	\begin{align}
			&\mathbf{P}(\frac{1}{T}\sum_{k=0}^{T-1}\sum_{i\in\mathcal{V}_{z_k}}\|\nabla_{x_i}f(x^k)\|^2\geq\epsilon)\nonumber\\
			&\leq\mathbf{P}(\frac{1}{T}\sum_{k=0}^{T-1}\sum_{i\in\mathcal{V}_{z_k}}\|\nabla_{x_i}f(x^k)\|^21_{\tau\geq T}\geq\epsilon)+\mathbf{P}(\tau< T)\nonumber\\
			&\leq\frac{1}{\epsilon}\mathbb{E}^k\left[\frac{1}{T}\sum_{k=0}^{T-1}\sum_{i\in\mathcal{V}_{z_k}}\|\nabla_{x_i}f(x^k)\|^21_{\tau\geq T}\right]+\mathbf{P}(\tau< T)\nonumber\\
			&\leq \frac{1}{\alpha}(\gamma+\frac{1}{\nu})+\frac{2+\nu\gamma}{\alpha}=\frac{1}{\alpha}(2+\gamma+\frac{1}{\nu}+\nu\gamma),\label{P>=epsilon}
	\end{align}
where the second inequality used Markov's inequality, and the last inequality used (\ref{gradnorm}), $T\geq \frac{2\alpha\nu f(x^0)}{\eta\epsilon}$ and $Z\leq \gamma\epsilon/\alpha$.

(ii). Without loss of generality, suppose that $T$ is divisible by $T_0$ (if not, the convergence accuracy $\epsilon$ can be re-selected from its small neighborhood such that $T=\lceil \frac{2\alpha \nu f(x^0)}{\eta\epsilon}\rceil$ is divisible by $T_0$.). Let $M=\frac{T}{T_0}$. For any $K\in\{0,...,M-1\}$, distinct steps $k, k'\in[KT_0,KT_0+T_0-1]$ and $x^k, x^{k'}\in\mathbb{X}$, we have 
\begin{align}
    &\|x^k-x^{k'}\|\leq \sum_{l=KT_0}^{KT_0+T_0-2}\|x^l-x^{l+1}\| \notag \\
    &\leq \sum_{l=KT_0}^{KT_0+T_0-2}\sum_{i\in\mathcal{V}_{z_l  }}\left(w_i^l\|x_i^l-x_i^{l_{prev}}\|+\eta\|g_i(\hat{x}_i^l,x_{-i}^l)\|\right)  \\ 
    &\leq \sum_{l=KT_0}^{KT_0+T_0-2}\sum_{i\in\mathcal{V}_{z_l}}\left(\frac{\bar{\epsilon}}{2(T_0-1)N_i}+\frac{\bar{\epsilon}}{2(T_0-1)N_0}\right)\leq \bar{\epsilon}\leq\rho_0,  \notag
\end{align}
where the third inequality utilized (\ref{extra conditions}), and $|\mathcal{V}_{z_l}|=N_i\leq N_0$. Then the smoothness of $f(x)$ at $x^k$ implies that 
\begin{equation}
	\|\nabla_{x}f(x^k)-\nabla_{x}f(x^{k'})\|\leq\phi_0\bar{\epsilon}.
\end{equation}
It follows that
	\begin{align}
	&\frac{1}{T_0}\|\nabla_xf(x^{k'})\|^2=\frac{1}{T_0}\sum_{j=1}^s\sum_{i\in\mathcal{V}_j}\|\nabla_{x_i}f(x^{k'})\|^2  \notag \\
	&\leq\frac{1}{T_0}\sum_{k=KT_0}^{(K+1)T_0-1}\sum_{i\in\mathcal{V}_{z_k}}\|\nabla_{x_i}f(x^{k'})\|^2 \notag\\
	&\leq \frac{1}{T_0}\sum_{k=KT_0}^{(K+1)T_0-1}\sum_{i\in\mathcal{V}_{z_k}}(2\|\nabla_{x_i}f(x^k)\|^2 \label{nablaxk'}\\
	&~~~~+ 2\|\nabla_{x_i}f(x^k)-\nabla_{x_i}f(x^{k'})\|^2)\notag\\
	&\leq \frac{1}{T_0}\sum_{k=KT_0}^{(K+1)T_0-1}\sum_{i\in\mathcal{V}_{z_k}}2\|\nabla_{x_i}f(x^k)\|^2+2\phi_0^2\bar{\epsilon}^2. \notag
  \end{align}
Then we have
\begin{align}
    &\frac{1}{T}\sum_{k=0}^{T-1}\|\nabla_xf(x^k)\|^2= \frac{T_0}{T}\sum_{K=0}^{M-1}\sum_{k'=KT_0}^{(K+1)T_0-1}\frac{1}{T_0}\|\nabla_xf(x^{k'})\|^2\nonumber\\
    &\leq \frac{T_0}{T}\sum_{K=0}^{M-1}\left[\sum_{k=KT_0}^{(K+1)T_0-1}\sum_{i\in\mathcal{V}_{z_k}}2\|\nabla_{x_i}f(x^k)\|^2+2T_0\phi_0^2\bar{\epsilon}^2\right]\nonumber\\
    &=\frac{T_0}{T}\left[\sum_{k=0}^{MT_0-1}\sum_{i\in\mathcal{V}_{z_k}}2\|\nabla_{x_i}f(x^k)\|^2+2MT_0\phi_0^2\bar{\epsilon}^2\right],
\end{align}
where the second inequality used (\ref{nablaxk'}).

Reusing (\ref{P>=epsilon}), the following holds with probability $1-\frac{1}{\alpha}(2+\gamma+\frac{1}{\nu}+\nu\gamma)$:
\begin{equation}
    \frac{1}{T}\sum_{k=0}^{T-1}\|\nabla_xf(x^k)\|^2< 2T_0(\epsilon+\phi_0^2\bar{\epsilon}^2).
\end{equation}
This completes the proof. \QEDA

	{\it Proof of Lemma \ref{le cov g as g}:} For notation simplicity, when the expectation is taken over all the random variables in a formula, we use $\mathbb{E}$ as the shorthand. Moreover, we use $\nabla_i$ to represent $\nabla_{x_i}$.	During the derivation, we will use the first-order approximation (assuming a sufficiently small $r$), i.e., 
	\begin{multline}
		h_i(x_i+ru_i,x_{-i},\xi) = h_i(x_i,x_{-i}, \xi) \\
		+ \nabla_i h_i(x_i,x_{-i}, \xi)ru_i+\mathcal{O}(r^2).
	\end{multline}
	
	For $g_l$, it holds that
		\begin{align}
		\mathbb{E}[g_lg_l^{\top}]&=\mathbb{E}[\frac{q_i}{r}h_i(x_i+ru_i,x_{-i},\xi)u_iu_i^{\top}\frac{q_i}{r}h_i(x_i+ru_i,x_{-i},\xi)] \notag\\
		&=\frac{q_i^2}{r^2}\mathbb{E}[h^2_i(x_i+ru_i,x_{-i},\xi)u_iu_i^{\top}] \notag\\
			&= \frac{q_i^2}{r^2} \mathbb{E}\left[(h_i(x, \xi) + ru^{\top}_i\nabla_i h_i(x, \xi)+\mathcal{O}(r^2))^2u_iu_i^{\top} \right] \notag\\
			&=\frac{q_i^2}{r^2}\mathbb{E}_{\xi}[h_i^2(x, \xi) +\mathcal{O}(r^2)]\mathbb{E}_{u_i}[u_iu_i^{\top}]\quad  \notag\\
			&= \frac{q_i}{r^2}\mathbb{E}[h_i^2(x, \xi)+\mathcal{O}(r^2)]I_{q_i} ,
		\end{align}
	where the fourth inequality is obtained by $\mathbb{E}(u_iu_i^{\top}u_i)=0$, and the last equality used $\mathbb{E}[u_iu_i^{\top}]=\frac{1}{q_i}I_{q_i}$ since $u_i\sim\text{Uni}(\mathbb{S}_{q_i-1})$.
	On the other hand,
		\begin{align}
			&\mathbb{E}[\frac{q_i}{r}h_i(x_i+ru_i,x_{-i},\xi)u_i] \notag\\
			&= \frac{q_i}{r}\mathbb{E}\left[(h_i(x, \xi)u_i + ru^{\top}\nabla_{x_i} h_i(x,\xi))u_i+\mathcal{O}(r^2)\right]\\
			&= \mathbb{E}[\nabla_{x_i}h_i(x,\xi)]+\mathcal{O}(r). \notag
		\end{align}
	As a result, $\Cov(g_l)$ is in (\ref{eq:cov1}).

	Similarly, 
	\begin{align}
		&\mathbb{E}[g_g] = \mathbb{E}[\frac{q}{r}h(x+rz,\xi)z_i]\nonumber\\
		&= \frac{q}{r}\mathbb{E}[h(x,\xi)z_i + rz^{\top}\nabla_{x_i} h(x,\xi)z_i+\mathcal{O}(r^2)]\nonumber\\
		&=q\mathbb{E}[z_iz^{\top}] \mathbb{E}[\nabla_xh(x,\xi)]+\mathcal{O}(r)\nonumber\\ &=\left[\mathbf{0}_{q_i\times q_1}~,\cdots,I_{q_i},\cdots,\mathbf{0}_{q_i\times q_N}\right] \mathbb{E}[\nabla_xh(x,\xi)]+\mathcal{O}(r)\nonumber\\
		&=\mathbb{E}[\nabla_{x_i}h(x,\xi)]+\mathcal{O}(r),
	\end{align}
	where the last equality used $\mathbb{E}[zz^{\top}]=\frac1qI_q$. Also
	\begin{align}
		\mathbb{E}[g_gg_g^{\top}] &= \frac{q^2}{r^2}\mathbb{E}\left[h(x+rz,\xi)z_iz_i^{\top}h(x+rz,\xi)\right]\nonumber\\
		&=\frac{q^2}{r^2}\mathbb{E}\left[(h(x,\xi)+rz^{\top}\nabla_{x} h(x,\xi)+\mathcal{O}(r^2))^2z_iz_i^{\top}\right] \nonumber\\ &=\frac{q}{r^2}\mathbb{E}[h^2(x,\xi)+\mathcal{O}(r^2)]I_{q_i}.
	\end{align}
	Thus, $\Cov(g_g)$ is in (\ref{eq:cov2}).
	\QEDA

	\subsection{Analysis for Application to Multi-Agent LQR}
	
	{\it Proof of Proposition \ref{pr LQR J_i}:} (i). Due to the boundedness of $x(0)$, there must exist $c_{lqr}>0$ such that for any $x(0)\sim\mathcal{D}$, it holds that $x(0)x^{\top}(0)\preceq c_{lqr}\mathbb{E}[x(0)x^{\top}(0)]=c_{lqr}\Sigma_x$. Let $Q_{i,K}^\infty=\sum_{t=0}^\infty\gamma^t(\mathcal{A}-\mathcal{B}K)^{t\top}(\hat{Q}_i+K^{\top}\hat{R}_iK)(\mathcal{A}-\mathcal{B}K)^{t}$. It follows that 
	\begin{equation}
		\begin{split}
			&H_i(K,x(0))=\langle Q_{i,K}^\infty,x(0)x^{\top}(0)\rangle\leq c_{lqr}\langle Q_{i,K}^\infty,\Sigma_x\rangle\\
			&=c_{lqr}\mathbb{E}_{x(0)\sim\mathcal{D}}[H_i(K,x(0))]=c_{lqr}J_i(K).
		\end{split}
	\end{equation}

	(ii). Given two control gains $K=(K_1^{\top},...,K_i^{\top},...,K_N^{\top})^{\top},$ $K'=(K_1^{\top},...,{K'_i}^{\top},...,K_N^{\top})^{\top}.$ Given the initial state $x(0)$, let $x(t)$ and $x'(t)$ be the resulting entire system state trajectories by implementing controllers $u=-Kx$ and $u=-K'x$, respectively. It holds for any $j\in\mathcal{V}$ that
	\begin{equation*}
x_j(t+1)=A_jx_j(t)+B_jK_jx(t)=A_jx_j(t)+B_j\tilde{K}_jx_{\mathcal{N}_S^j}(t). 
	\end{equation*}
	Note that for all $j\in\mathcal{V}\setminus\mathcal{N}_L^i$, it must hold that $\mathcal{N}_C^j\cap\mathcal{V}_S^i=\varnothing$. From the definition of $\mathcal{V}_S^i$, we have  $x_j(t)=x'_j(t)$ and $x_{\mathcal{N}_S^j}(t)=x'_{\mathcal{N}_S^j}(t)$ for all $j\in\mathcal{V}\setminus\mathcal{N}_L^i$. It follows that $x^{\top}(Q-\hat{Q}_i)x=x'^{\top}(Q-\hat{Q}_i)x'$, and for any $j\in\mathcal{V}\setminus\mathcal{N}_L^i$, we have	$$x^{\top}K_j^{\top}R_jK_jx=x_{\mathcal{N}_S^j}^{\top}\tilde{K}_j^{\top}R_j\tilde{K}_jx_{\mathcal{N}_S^j}= x^{'\top}_{\mathcal{N}_S^j}\tilde{K}_j^{\top}R_j\tilde{K}_jx'_{\mathcal{N}_S^j}.$$
	Therefore,
		\begin{align}
			&J(K)-J(K')\notag\\
			&=\mathbb{E}\left[\sum_{t=0}^{\infty}x^{\top}(Q+K^{\top}RK)x\right]-\mathbb{E}\left[\sum_{t=0}^{\infty}{x'}^{\top}(Q+K^{\top}RK)x'\right] \notag\\
			&=\mathbb{E}\left[\sum_{t=0}^{\infty}x^{\top}(\hat{Q}_i+\sum_{j\in\mathcal{N}_L^i}K_j^{\top}R_jK_j)x\right]\notag\\
			&~~~~-\mathbb{E}\left[\sum_{t=0}^{\infty}{x'}^{T}(\hat{Q}_i+\sum_{j\in\mathcal{N}_L^i}{K'}_j^{\top}R_jK'_j)x'\right] \notag \\
			&=J_i(K)-J_i(K').
		\end{align}
	This means that any perturbation on $K_i$ results in the same difference on $J(K)$ and $J_i(K)$. The proof is completed.
	\QEDA

	{\it Proof of Lemma \ref{le LQR g as}:} 
	Note that
 \begin{small}
	\begin{align}
	        &H_i(K^{i,k},x(0))-H_{i,T_J}(K^{i,k},x(0))\nonumber\\
	        &=\sum_{t=T_J}^\infty\gamma^t x^{\top}(t)(\hat{Q}_i+K^{i,k\top}\hat{R}_iK^{i,k})x(t)\nonumber\\
	        &=\sum_{t=T_J}^\infty y^\top(t)(\hat{Q}_i+K^{i,k\top}\hat{R}_iK^{i,k})y(t)    \\
	        &=x^\top(0)\left((\sqrt{\gamma}(\mathcal{A}-\mathcal{B}K^{i,k}))^{T_J}\right)^\top P_i\left(\sqrt{\gamma}(\mathcal{A}-\mathcal{B}K^{i,k})\right)^{T_J}x(0)\nonumber\\
	        &\leq \|x(0)\|^2 \text{trace}(P_i)\|\left(\sqrt{\gamma}(\mathcal{A}-\mathcal{B}K^{i,k})\right)^{T_J}\|^2\nonumber
	\end{align}
\end{small}
	where $y(t)=\gamma^{t/2}x(t)$, the third equality comes from the fact $y(t+1)=\sqrt{\gamma}(\mathcal{A}-\mathcal{B}K)y(t)$, and $P_i$ is the solution to
	\begin{equation*}
	    P_i=\hat{Q}_i+K^{i,k\top} \hat{R}_iK^{i,k}+\gamma (\mathcal{A}-\mathcal{B}K^{i,k})^\top P_i(\mathcal{A}-\mathcal{B}K^{i,k}).
	\end{equation*}
Since $\mathbb{E}[x^\top(0)P_ix(0)]=J_i(K^{i,k})$, we have $\text{trace}(P_i)\leq \frac{J_i(K^{i,k})}{\lambda_{\min}(\Sigma_x)}\leq\frac{\alpha_iJ_i(K^0)}{\lambda_{\min}(\Sigma_x)}$. Together with (\ref{C_K}) and $K^k\in\mathbb{K}_\alpha$, we have
 \begin{equation}\label{Vi-Vf}
	    H_i(K^{i,k},x(0))-H_{i,T_J}(K^{i,k},x(0))\leq \varpi\alpha_iJ_i(K^0)C_0^2\kappa_0^{2T_J},
	\end{equation}
where $\varpi=\frac{\lambda_{\max}(x(0)x^\top(0))}{\lambda_{\min}(\Sigma_x)}$. 
	It follows from~\eqref{T_J bound} that
 \begin{equation*}\label{eq:2TJ}
	   2T_J\geq \frac{\log \left(\varpi\alpha_iJ_i(K^0)C_0^2/\epsilon'\right)}{\log(1/\kappa_0)}=\log_{\kappa_0}\frac{\epsilon'}{\left (\varpi\alpha_iJ_i(K^0)C_0^2\right)},
	\end{equation*}
where the inequality used the fact $1-\kappa_0\leq \log(1/\kappa_0)$.
	Substituting the lower bound of $2T_J$ into~\eqref{Vi-Vf} yields $ H_i(K^{i,k},x(0))-H_{i,T_J}(K^{i,k},x(0))\leq\epsilon'$. 	

	Next we  prove (\ref{LQR g error}). For any $i\in\mathcal{V}$, we have
	\begin{align}
			&\|\mathbb{E}[\mathbf{G}_i(k)-\mathbf{G}_i^{T_J}(k)]\|\nonumber\\
			&\leq|\frac{q_i}{r_i}(H_i(K^{i,k},x(0))-H_{i,T_J}(K^{i,k},x(0)))| \cdot\max_{\mathbf{D}_i\in\mathbb{S}_{q_i-1}}\|\mathbf{D}_i\|\nonumber\\
			&\leq \frac{q_i}{r_i}\epsilon'.
	\end{align}
	According to Lemma \ref{le fhat error}, when $r_i\leq\rho_0\leq\beta_K$, we have
	\begin{equation}
		\|\mathbb{E}[\mathbf{G}_i(k)]-\nabla_{\mathbf{K}_i}J(K^{i,k})\|\leq \phi_0r_i.
	\end{equation}
	Using the triangular inequality yields (\ref{LQR g error}).
	
	Next we bound $\|\mathbf{G}^{T_J}_i(k)\|$:
	\begin{equation}
		\begin{split}
			\|\mathbf{G}_i^{T_J}(k)\|&\leq \frac{q_i}{r_i}(H_i(K^{i,k},x(0))+\epsilon')\\
			&\leq\frac{q_i}{r_i}(c_{lqr}J_{i}(K^{i,k})+\epsilon')\\
			&\leq \frac{q_i}{r_i}\left[c_{lqr}(J_i(K^{k})+\lambda_0\rho_0)+\epsilon'\right]
			\\
			&\leq \frac{q_i}{r_i}\left[c_{lqr}(\alpha_i  J_i(K^0)+\lambda_0\rho_0)+\epsilon'\right],
		\end{split}
	\end{equation}
where the first inequality used the first statement in Proposition \ref{pr LQR J_i}, the second inequality used 
 $$\|\mathbf{K}^{i,k}-\mathbf{K}^k\|\leq\|\hat{\mathbf{K}}_i^k-\mathbf{K}_i^k\|+r_i=w_i\|\mathbf{K}_i^k-\mathbf{K}_i^{k_{prev}}\|+r_i\leq \rho_0,$$
the third inequality used  $J_i(K^k)\leq \alpha_i J_i(K^0)$.
	\QEDA

	
	

	\bibliography{References.bib} 
	\bibliographystyle{IEEEtran} 

%

\vskip -2\baselineskip plus -1fil
\begin{IEEEbiography}[{\includegraphics[width=1in,height=1.25in,clip,keepaspectratio]{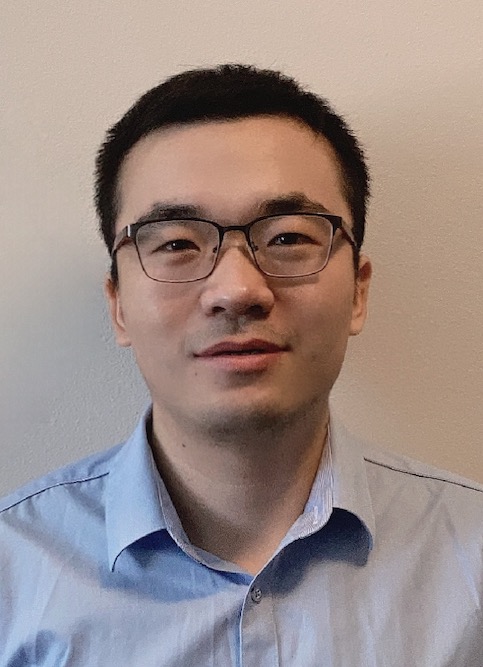}}]
{Gangshan Jing} received the Ph.D. degree in Control Theory and Control Engineering from Xidian University, Xi'an, China, in 2018. From 2016 to 2017, he was a research assistant at Hong Kong Polytechnic University, Hong Kong. From 2018-2021, he was a postdoctoral researcher at Ohio State University, USA and North Carolina State University, USA, respectively. Since Dec. 2021, he has been a faculty member at Chongqing University, China. His research interests include control, optimization, and machine learning for network systems.
\end{IEEEbiography} 
\vskip -2\baselineskip plus -1fil
\begin{IEEEbiography}   
[{\includegraphics[width=1in,height=1.25in,clip,keepaspectratio]{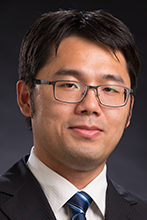}}]
{He Bai} received his Ph.D. degree in Electrical Engineering from Rensselaer Polytechnic Institute, Troy, NY, in 2009. From 2009 to 2010, he was a postdoctoral researcher at Northwestern University, Evanston, IL. From 2010 to 2015, he was a Senior Research and Development Scientist at UtopiaCompression Corporation, Los Angeles, CA. In 2015, he joined the School of Mechanical and Aerospace Engineering at Oklahoma State University, Stillwater, OK, where he is currently an associate professor. His research interests include distributed estimation, control and learning, reinforcement learning, nonlinear control, and robotics. 
\end{IEEEbiography}
\vskip -2\baselineskip plus -1fil
\begin{IEEEbiography}
[{\includegraphics[width=1in,height=1.25in,clip,keepaspectratio]{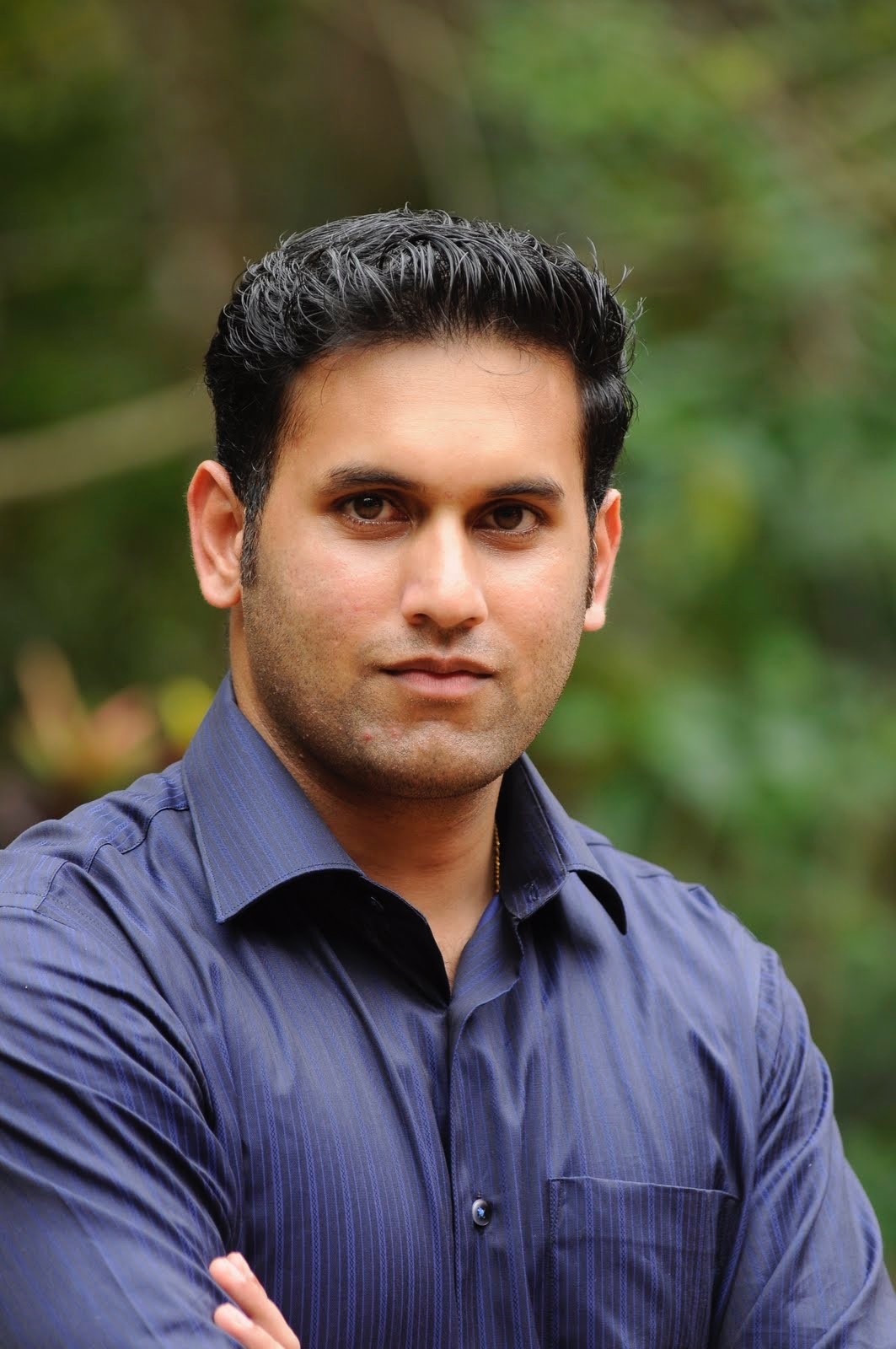}}]
{Jemin George} received his M.S. (’07), and Ph.D. (’10) in Aerospace Engineering from the State University of New York at Buffalo. Prior to joining ARL in 2010, he worked at the U.S. Air Force Research Laboratory’s Space Vehicles Directorate and the National Aeronautics, and Space Administration's Langley Aerospace Research Center. From 2014-2017, he was a Visiting Scholar at the Northwestern University, Evanston, IL. His principal research interests include distributed learning, stochastic systems, control theory, nonlinear estimation/filtering, networked sensing and information fusion. 
\end{IEEEbiography} 
\vskip -2\baselineskip plus -1fil
\begin{IEEEbiography}
[{\includegraphics[width=1in,height=1.25in,clip,keepaspectratio]{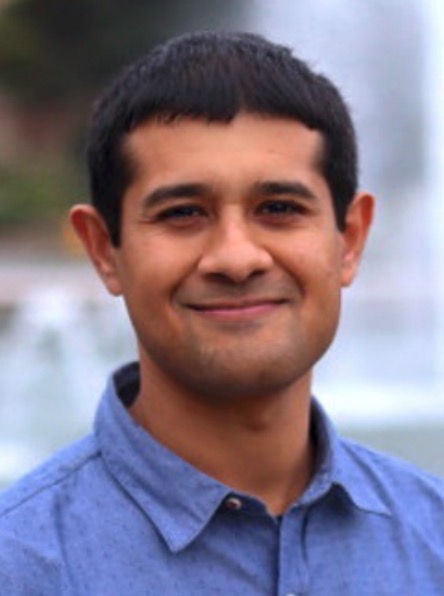}}]
{Aranya Chakrabortty} received the Ph.D. degree in
Electrical Engineering from Rensselaer Polytechnic Institute, NY in 2008. From 2008 to 2009 he was a postdoctoral research associate at University of
Washington, Seattle, WA. From 2009 to 2010 he was an assistant professor at Texas Tech University, Lubbock, TX. Since 2010 he has joined the Electrical and Computer Engineering department at North Carolina State University, Raleigh, NC, where he is currently a Professor. His research interests are in all branches of control theory with applications to electric power systems. He received the NSF CAREER award in 2011.
\end{IEEEbiography} 
\vskip -2\baselineskip plus -1fil
\begin{IEEEbiography}
[{\includegraphics[width=1in,height=1.25in,clip,keepaspectratio]{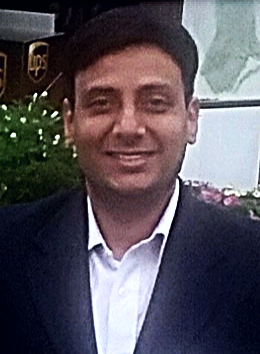}}]
{Piyush Sharma} received his M.S. and Ph.D. degrees in Applied Mathematics from the University of Puerto Rico in 2011 and Delaware State University in 2016, respectively. He has government and industry work experiences. Currently, he is with the U.S. Army as an AI Coordinator at ATEC, earlier a Computer Scientist at DEVCOM ARL. Prior to joining ARL, he worked for Infosys’ Data Analytics Unit (DNA), where his role was a Senior Associate Data Scientist. He has a core data science experience and has been responsible for Thought Leadership and solving stakeholders’ problems, communicating results, and methodologies with clients. His principal research includes multiagent reinforcement learning, multimodality learning in multiagent systems, decision-making, and Internet of Battlefield Things.
\end{IEEEbiography} 	
\end{document}